\documentclass{emulateapj} \usepackage{times}

\usepackage{color} 
\usepackage{epsfig}

\newcommand{\dc}{$D$-criterion }
\newcommand{\etal}{{et~al.}} 
\newcommand{\eg}{{\it e.g.}}
\newcommand{\ie}{{\it i.e.}} 

\shorttitle{Identifying NEO Families} 
\shortauthors{Fu \etal} 
\journalinfo{accepted for publication in Icarus}
\submitted{accepted for publication in Icarus}

\begin{document}

\title{Identifying Near Earth Object Families}

\author{Hai Fu, Robert Jedicke} 
\affil{Institute for Astronomy, University of Hawaii \\
2680 Woodlawn Drive, Honolulu, HI, 96822}
\email{fu@ifa.hawaii.edu, jedicke@ifa.hawaii.edu}

\vskip 1.0cm

\author{Daniel D. Durda} 
\affil{Southwest Research Institute \\
1050 Walnut Street Suite 400, Boulder, CO, 80302}
\email{durda@boulder.swri.edu}
 
\vskip 1.0cm

\and

\author{Ronald Fevig, James V. Scotti} 
\affil{Lunar and Planetary Lab, University of Arizona \\
P.O. Box 210092, Tucson, AZ, 85721-0092}
\email{fevig@lpl.arizona.edu, jscotti@lpl.arizona.edu}

\clearpage

\vskip 2.0cm

\begin{abstract}

The study of asteroid families has provided tremendous insight into
the forces that sculpted the main belt and continue to drive the
collisional and dynamical evolution of asteroids.  The identification of
asteroid families within the NEO population could provide a similar
boon to studies of their formation and interiors. 
In this study we examine the purported identification of NEO
families by \citet{dru00} and conclude that it is unlikely that they
are anything more than random fluctuations in the distribution of NEO
osculating orbital elements.  We arrive at this conclusion after
examining the expected formation rate of NEO families, the
identification of NEO groups in synthetic populations that contain no
genetically related NEOs, the orbital evolution of the largest
association identified by \citet{dru00}, and the decoherence of
synthetic NEO families intended to reproduce the observed members of
the same association.  These studies allowed us to identify a new
criterion that can be used to 
select real NEO families for further study in future analyses, based on the ratio of the number of $pairs$ and the size of
$strings$ to the number of objects in 
an identified $association$.

\end{abstract}


\section{Introduction}
\label{sec.intro}

The discovery of a group of Near Earth Objects (NEO) that all derive
from the products of a catastrophically disrupted NEO parent body
would be an important step in understanding their structure and
dynamical evolution.  It would be equivalent to the information
afforded a geologist in cracking open a rock with a hammer.  Studies
of the chips off the old block of a larger asteroid will provide
interesting insights, just as the study of similarly formed main belt
(MB) asteroid families has provided critical information to the study
of those asteroids \citep[\eg,][]{cel04,zap02}.  The difference is that
NEO families will have a much faster dynamical evolution such that
they are detectable over much shorter timescales as they diffuse
into the orbital element space of the background objects.  They
also sample a much different range of parent asteroid sizes since the
known sample of NEOs are much smaller than the known sample of MB objects.

The time scale of catastrophic collisions between MB asteroids that
create asteroid families is short compared to the age of the Solar
System.  In catastrophic collisions, the colliding bodies completely
shatter into smaller fragments, some of which are dispersed and then
re-accumulate into objects on independent but similar orbits
\citep{dav02}. The first groups of asteroids with nearly the same
orbital elements were discovered by \citet{hir18} and these groups are
now known as {\it Hirayama families}.  The members of each family
usually share similar spectral properties \citep{cel02}, further
suggesting a common origin in a single parent body that has undergone
a catastrophic collision. We consider the members of this kind of
asteroid family ``genetically" related since they share the same parent
body. At this time, more than 50 significant families have been
tabulated in the MB \citep[\eg,][]{nes05}.

MB families are identified by searching for enhancements in the space
of {\it proper} orbital elements ($a_p, e_p, i_p$). A precise definition
of the proper elements \citep[\eg,][]{kne02} is beyond the scope of
this paper but it is sufficient here to think of them as long-term
average orbital elements.  This method is not readily applicable to
NEOs for two main reasons: (1) the calculation of proper elements for
NEOs is problematic due to the extremely chaotic evolution of NEO
orbits and, (2) the limited population (3,319 objects as of 2005
Apr. 28) of known NEOs are spread over a much larger orbital element
space than MB asteroids. Therefore, while asteroid families have long
been recognized in the MB, the situation remains unclear regarding the
existence of genetically related families other than meteor streams in
the NEO population.

We have identified five possible mechanisms for the production of
genetic NEO families: 
\begin{enumerate}
\item NEO-MB asteroid collisions, 
\item tidal disruption of NEOs, 
\item intra-NEO collisions, 
\item a MB family producing event near a MB resonance followed by
rapid dynamical evolution of some of the members into NEO orbits, and
\item spontaneous NEO disruption.  
\end{enumerate}
Among the five, we consider the first two to be the
most likely method for producing NEO families.  Collisions between
NEOs are expected to be rare \citep{bot96} since the density of NEOs
interior to the MB is miniscule compared to the density of asteroids
in the MB. The fourth mechanism is unlikely because there is only a
small fraction of MB orbital element space near resonances capable of
transporting the residue onto NEO orbits.  Furthermore, the chaotic
dynamical evolution during the transportation of the fragments to
near-Earth orbits makes it unlikely that such a family could be
identified as such based on their osculating elements. The last
mechanism is reserved for objects with high volatile content,
presumably comets that have worked their way into NEO orbits, and is
probably the way meteor streams \citep[\eg,][]{dru81} are created.
Thus, meteor streams are genetic NEO families but the members are
orders of magnitude smaller than the smallest detectable NEOs.

The number of catastrophic NEO collisions per year that may produce families
due to collisions on targets up to absolute magnitude $H_{max}$ is given by:
\begin{equation}
N_{family}(<H_{max}) = \int_{-\infty}^{H_{max}} n(H) \;  p_C(H) \; dH
\end{equation} 
where $n(H)$ is the differential number distribution of NEOs, and
$p_C(H)$ is the annual probability of catastrophic collision, both
given as a function of absolute magnitude.  Using the NEO
size-frequency distribution (SFD) of \citet{bot02a} and the collision
probability {\it for MB asteroids} from Fig. 14 of \citet{bot05}, we
estimate that during the typical $10^6$ year dynamical lifetime of
NEOs \citep{mor02,gla97} $\lll$ 1/$\sim$8/$\sim$1000 NEOs larger than
10/1/0.1 km diameter are catastrophically disrupted.  However, the
catastrophic collision probability for NEOs is dramatically smaller
than that of MB asteroids due to their much lower space-density
\citep{bot96}.  Once an NEO is dynamically decoupled from the MB the
collision probability drops even further.  Thus, we consider it
unlikely that any NEOs $>$1 km diameter have suffered a
family-producing event while the number of NEOs $>$100 m diameter that
have produced families is small.

An asteroid family produced in the disruption of a target would
produce a large number of smaller asteroids.  The SFD of the
re-accumulated fragments may be determined through numerical modelling
\citep[\eg,][]{mic02,dur05} or through studies of the SFD of known MB
asteroid families \citep[\eg,][]{tan99}. The SFDs of the families are
usually described by a power-law in the diameter $D$, $N(>D) \propto
D^{-p}$, where $2 \lesssim p \lesssim 5$ for real MB families (\eg,
Eunomia, Flora \& Koronis) though it appears \citep{mor03} that the
SFD of observed families becomes shallower for $H>15$ ($D\lesssim
5$~km).  The SFD of simulated new asteroid families displays a rich
morphology \citep{mic02,dur05} with a wide range of slopes that is
difficult to characterize with a single power-law over the entire
range of sizes but they are qualitatively similar to the
observed SFD.

For the purpose of quickly estimating the expected number of known
fragments due to a NEO family producing event, we assume that the
fragments have a SFD similar to the known families with $2 \lesssim p
\lesssim 5$.  In a barely catastrophic collision the largest fragment
(LF) has half the mass of the target.  Since we know that $M \propto V
\propto D^3$ and that the diameter of an asteroid is related to its
absolute magnitude by $D \propto 10^{-H/5}$, then it is simple to show
that $H_{LF} = H_T + 0.5$ where $H_T$ is the absolute magnitude of the
target.  By definition, $N(\le H_{LF}) = 1$, and thus $N(\le H) =
10^{\alpha(H-H_T-0.5)}$ so that the differential HFD for the fragments
must be:
\begin{equation}
n(H) \sim 2.3 \; \alpha \; 10^{\alpha(H-H_T-0.5)}
\end{equation}
with $0.4 \lesssim \alpha \lesssim 1.0$ and $H_T$ being the absolute
magnitude of the target.

If the actual and known population in a volume $dadedidH$ are
represented by $dN$ and $dn$ respectively, then $dn =
\epsilon(a,e,i,H) dN$ where $\epsilon$ is the observational completion
of NEOs with orbital elements in the range $a \rightarrow a + da$, $e
\rightarrow e + de$, $i \rightarrow i + di$ and $H \rightarrow H + dH$.  The
number of known fragments from the collision is then
\begin{equation}
N_{known} = \int_{-\infty}^{+\infty} n(H) \; \epsilon(H) \; dH
\end{equation}
where $\epsilon(H) \sim \epsilon(a^\prime,e^\prime,i^\prime,H)$ and
$(a^\prime,e^\prime,i^\prime)$ are the mean orbital elements
corresponding to a particular family.

Using reasonable estimates for $\epsilon(H)$ and the full range of
possible slopes for the SFD, we find that the number of detectable
fragments in a NEO family produced in the collision with a 1 km
diameter target varies from essentially zero to many tens of thousands
(if the family has the most steep SFD).  This has
tremendous implications for the formation of hazardous NEO families
and spikes in the collision probability with the Earth.  However, as
argued above, the catastrophic disruption of a 1 km diameter NEO is
highly unlikely.  In the case of a 100m diameter NEO target the number
of observable fragments is essentially zero.  The fragments would be
very small and would quickly begin to evolve apart dynamically and be
strongly affected by non-gravitational forces \citep[\eg, Yarkovsky
forces: see][for a review]{bot02b}.  Our ability to recognize a
genetically related NEO family will depend on the efficiency of the
employed algorithm and the time scale for decoherence of the members"
orbits.  We will explore both these factors in this work.

Thus, it would seem unlikely that observable NEO families can be
produced and identified in the known NEO population.  That being said,
the opportunity of finding an NEO family would be important to
asteroid physical studies and would also allow us to refine our
understanding of the catastrophic disruption process and statistics.
In a similar vein, \citet{pau05} have studied the decoherence of
``meteorite streams" to determine if time-correlated meteorite falls
exist.

\citet{dru91,dru00} made the first attempts at searching
for groupings in the known NEO population. His study was based on a
\dc analysis (orbit similarity) of {\it osculating} orbital elements.
Osculating elements are the instantaneous 2-body (Sun+object) orbital
elements for an object.  He discovered 14 associations, 8 strings and
7 pairs out of a sample of 708 NEOs. Briefly, associations are density
enhancements above the local background in a four-dimensional \dc
space (the values in each of the four dimensions are functions of an
object"s five {\it osculating} elements not including the mean
anomaly). Strings are sets of NEOs in which every member is
``connected" to at least one other member with \dc less than a
specified threshold value, and pairs are two objects that have orbits
that are statistically determined to be improbably close. These
groupings contain a total of 155 NEOs or $\sim$ 22\% of the entire
sample known at the time of Drummond's study.

\placetable{tab.A1Association}

At this time, 56 objects in his groupings (36\%) have been taxonomically
classified\footnote{Data collected from the PDS Asteroid/Dust Subnode,
http://www.psi.edu/astsubnode.html} but there is no obvious correlation
between the taxonomic types in his groupings. Moreover, there are
cases where very different taxonomic types exist within the same
grouping --- for instance, both S-type and F-type NEOs exist in the A1
association (the largest association identified by \citet{dru00}, see
Table~\ref{tab.A1Association}).  In MB families there is a high degree of
correlation between the colors or taxonomic types of objects associated
with each family \citep[\eg,][]{nes05, cel02}.   The fact that there exists heterogeneity in
Drummond's (2000) NEO associations may be used to argue that the groups
are not genetically related families.  Alternatively, it may be argued
that the groups represent the fragments of an inhomogeneous parent
body.  The latter explanation is probably unlikely given that the NEOs
are themselves probably fragments from collisions that took place
within the MB.

In this paper we test the significance of the purported NEO families
through various numerical simulations. First, we briefly describe (\S\ref{sec.NEOFamilyIdentification}) the
analysis technique employed by \citet{dru91,dru00}. In
\S\ref{sec.StatisticalSignificanceOfNEOGroups} we discuss the statistical significance of the
detections by attempting to search for groupings in synthetic NEO
populations where no genetic families exist. We then demonstrate the decoherence of synthetic A1-like families as they
evolve under only the effect of gravitational forces in \S\ref{sec.DecoherenceOfSyntheticA1Family} and study the
dynamical evolution of the actual A1 association in \S\ref{sec.ActualA1Evolution}.  These studies
allowed us to develop a technique for identifying real families within
the NEO population as described in
\S\ref{sec.IdentifyingActualGeneticFamilies}. We discuss our 
main results in \S\ref{sec.Discussion} and close with a summary in 
\S\ref{sec.Conclusion}.

\section{NEO Family identification}
\label{sec.NEOFamilyIdentification}


Several versions of the \dc have been introduced
\citep[\eg,][]{sou63,dru81,jop93,val99} to quantify orbital
similarities.
The dimensionless \dc used in \citet{dru00} and adopted in this
study was defined by \citet{sou63} as: 
\begin{equation} 
D = \sqrt{d_{1}^{2}+d_{2}^{2}+d_{3}^{2}+d_{4}^{2}}, 
\end{equation} 
where
\begin{displaymath} 
d_{1}^{2} = ({q_1 - q_2 \over AU})^2;
\indent d_{2}^{2} = (e_1-e_2)^2; 
\end{displaymath}
\begin{equation}
\indent d_{3}^{2} = [2 \sin(I/2)]^2;
\indent d_{4}^{2} = [(e_1+e_2) \sin(\Pi/2)]^2 
\end{equation} 
and 
\begin{equation}
I =\arccos[\cos i_1 \cos i_2 + \sin i_1 \sin i_2 \cos(\Omega_1-\Omega_2)]
\end{equation}
\begin{equation}
\label{eqn:Pi} 
\Pi = \omega_1 - \omega_2 + 2\arcsin[\cos \frac{i_1+i_2}{2} \sin \frac{\Omega_1-\Omega_2}{2} \sec \frac{I}{2}].
\end{equation} 
The subscripts 1 and 2 refer to the two orbits being
compared, and the sign of the $\arcsin$ term of Eq. \ref{eqn:Pi}
changes when $\mid \Omega_1-\Omega_2\mid > 180^{\circ}$.  $q=a(1-e)$ is the
perihelion distance, $e$ the eccentricity, $i$ the inclination, $\omega$
the argument of perihelion, and $\Omega$ the longitude of the ascending node. 

The \dc is a dimensionless distance metric in a four-dimensional
space. The volume search method introduced by \citet{dru00} to
identify associations
searches for density enhancements above the local background around every
object in the sample in this four-dimensional space. It estimates the
local background level by fitting a uniform local background and a
Gaussian association (a gamma distribution) simultaneously to the
cumulative distribution of $D^2$ around each orbit\footnote{We used {\it
Pikaia} 1.2 \citep{cha02}, a FORTRAN version of a genetic search
algorithm, as our non-linear least square fitting code to find the global
minimum $\chi^2$.}, and a candidate group is identified if the
association plus background model gives a better fit to the data than the
background-only model. Those asteroids with \dc less than the
value at which the association density equals the background density are
gathered to form a mean orbit around which the cumulative distribution of
$D^2$ is fit for the two models again. The procedure iterates until it
converges on a stable group of objects. The groupings detected in this way
were termed {\it associations} by \citet{dru00}.  


We also searched for {\it pairs} and {\it strings}
 based on the \dc in a manner similar to that employed by \citet{dru00}. A couple of NEOs are a pair if
the \dc between the two is below a cutoff $D_{pair}$. Similarly,
a string of NEOs is a set of objects in which every member is connected
to at least one other with a $D < D_{string}$. In \citet{dru00}, a
cutoff of $D_{string} = 0.115$ was chosen under the guidance of the
results from the volume search. We set a more stringent threshold,
$D_{string} = D_{pair} = 0.1$, throughout this paper for the reason discussed below (\S\ref{sec.IdentifyingActualGeneticFamilies}). 

\begin{figure} 
\centerline{\epsfig{file=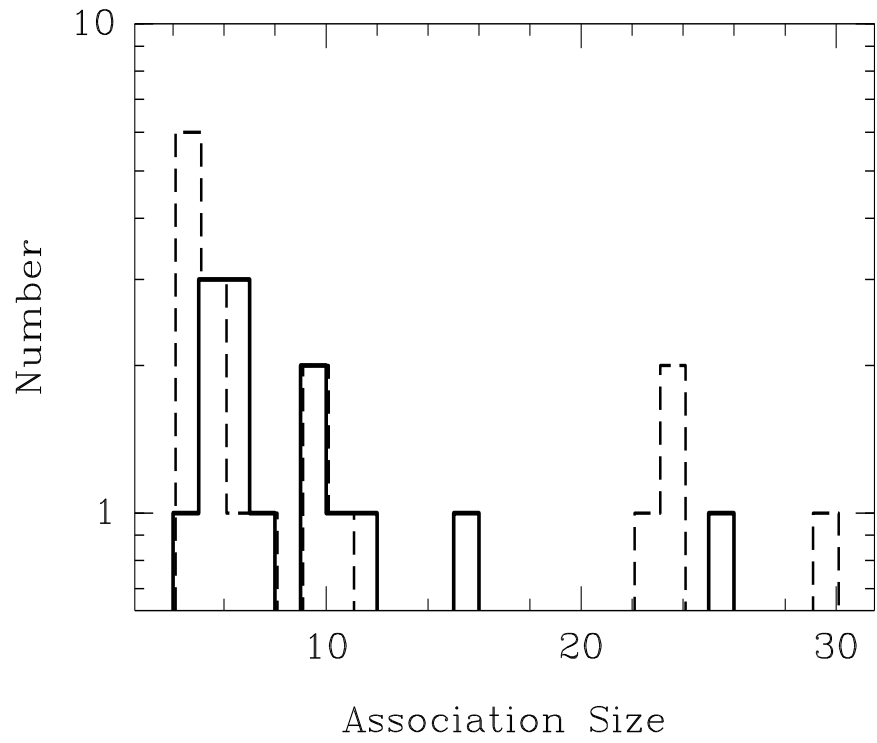, angle=0,
width=5in, bbllx=50pt, bblly=200pt, bburx=410pt, bbury=510pt} \\}
\caption{Size distribution of associations published in \citet{dru00}
(solid line) and the associations detected by the volume search code
of this analysis (dashed line).}\label{fig.Dru.vs.us} 
\end{figure}

\placefigure{fig.Dru.vs.us}
\placetable{tab.LargeGroupComparison}

In Drummond's (2000) study, the search for groupings was performed on a
sample of 708 NEOs (through 1999~HF1) listed on the Minor Planet Center
(MPC)
homepage as of 1999 April 22. As a test of our code, we attempted to
reproduce Drummond's associations by running our volume search code
on the same sample. Like \citet{dru00} we often identified 
overlapping associations in
which one association was partially or wholly contained within
another. We eliminated the {\it smaller} of any two associations that
shared more than 66\% of its members with the larger group.
Fig.~\ref{fig.Dru.vs.us} shows that our result is
slightly different from \citet{dru00} although there is rough
agreement in the distribution of the sizes of the associations.
Table~\ref{tab.LargeGroupComparison} compares the sizes and overlap
between the four largest associations found in this work and in
\citet{dru00}.  There is clearly a great deal of overlap between the
groupings found in both works but the relationship is not perfect.

We believe there are three main reasons for the disagreement: (1) The
orbital elements for the 708 NEOs are no longer identical to what they
were five years ago.  Most of these objects have much better orbits
now due to new observations provided to the MPC in the intervening
time period; (2) The volume search technique is very sensitive to the
input orbital elements.  Convergence of the iterative technique for
identifying associations can alter membership by a few objects with
changes in the \dc of only 0.01 (If all the difference between
two orbits is in either $a$ or $e$ then $\Delta a = 0.01$ or $\Delta e
= 0.01$ respectively). 
Reproducing the osculating elements of the 708 NEOs
used by Drummond would be a monumental task involving extracting only
those detections known to be associated with each object at the time
that \citet{dru00} obtained the orbital element data and then
determining the orbit; (3) The technique often identified
overlapping associations that shared many members. As described above, when
the overlap between two groups exceeded 66\% the smaller one was
discarded which biases our method towards larger groups.

Since we believe that our
technique is demonstrably identical to \citet{dru00} given the result
shown in Fig.~\ref{fig.Dru.vs.us} we proceed with the use of the
contemporary osculating elements for all 708 NEOs.

\section{Statistical significance of NEO groups} 
\label{sec.StatisticalSignificanceOfNEOGroups}


\citet{dru00} tested his NEO grouping software on synthetic populations
of NEOs but did not, apparently, take into account observational
selection effects when creating those populations.  This is not
surprising considering the difficulties involved in creating a realistic
NEO population and then simulating the surveying capability of the many
observatory programs that contributed to the 708 NEOs in his sample.
Drummond tested the veracity of his groupings by performing the
volume search on two sets of randomized orbital elements and comparing
them to the results for the real set. He concluded that 50 to 70\% of
the groupings he had found might not be real. 

Since that time, a good four-dimensional orbital element and absolute
magnitude model of the NEOs has been developed by \citet{bot02a} and good
matches have been obtained between simulated NEO surveys and the known
objects \citep[\eg,][]{ray04,jed03}.  

\begin{figure} 
\epsfig{file=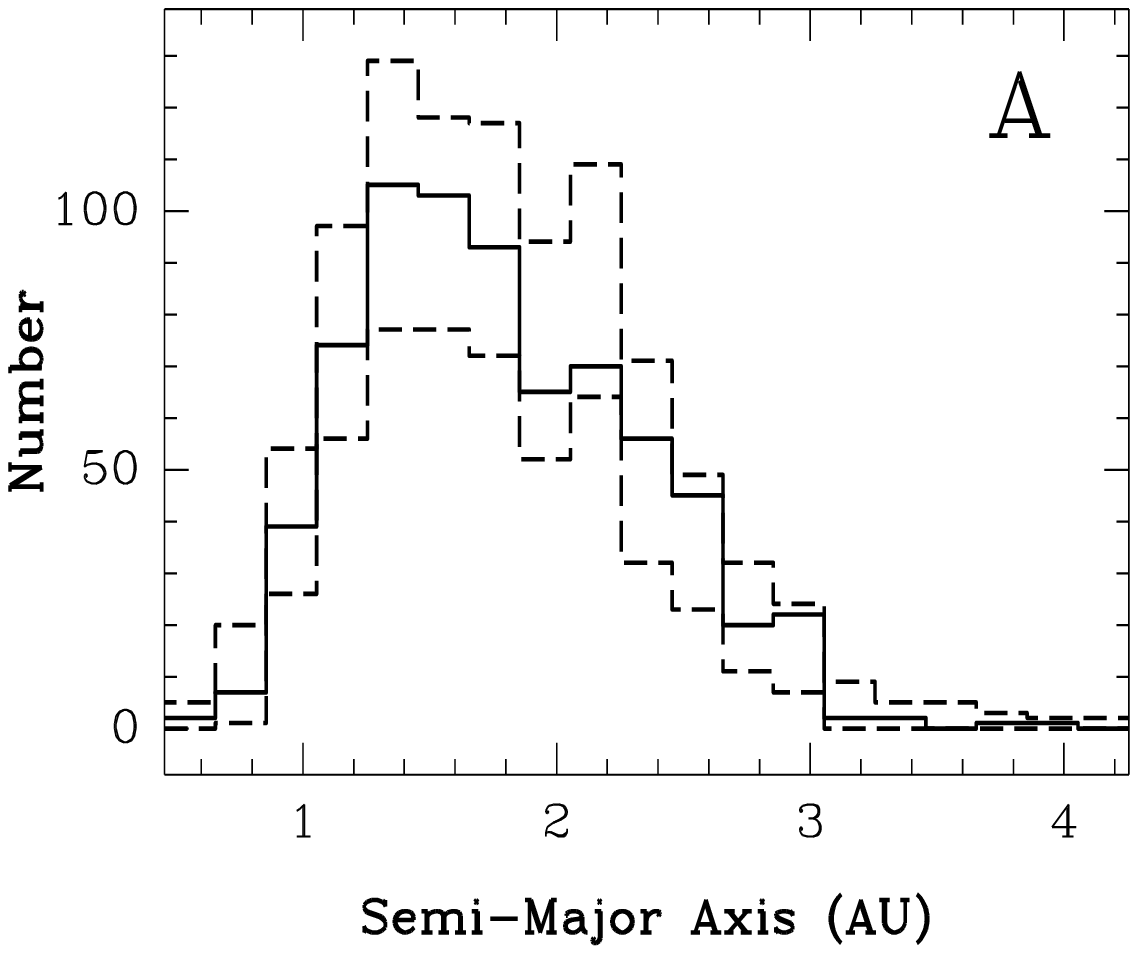, angle=0, width=4.0in,
bbllx=120pt, bblly=200pt, bburx=480pt, bbury=510pt}
\epsfig{file=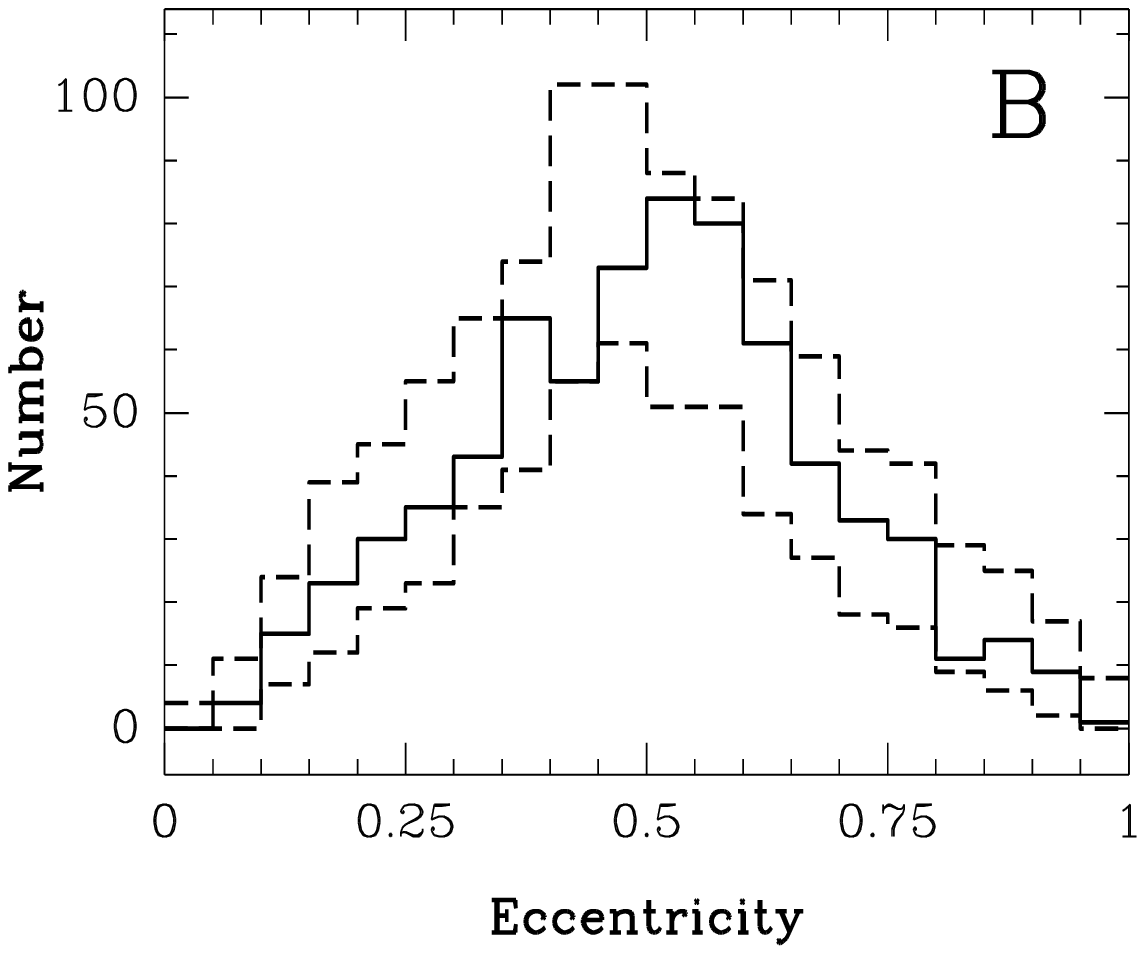, angle=0, width=4.0in, 
bbllx=140pt,bblly=200pt, bburx=500pt, bbury=510pt} \\ 
\epsfig{file=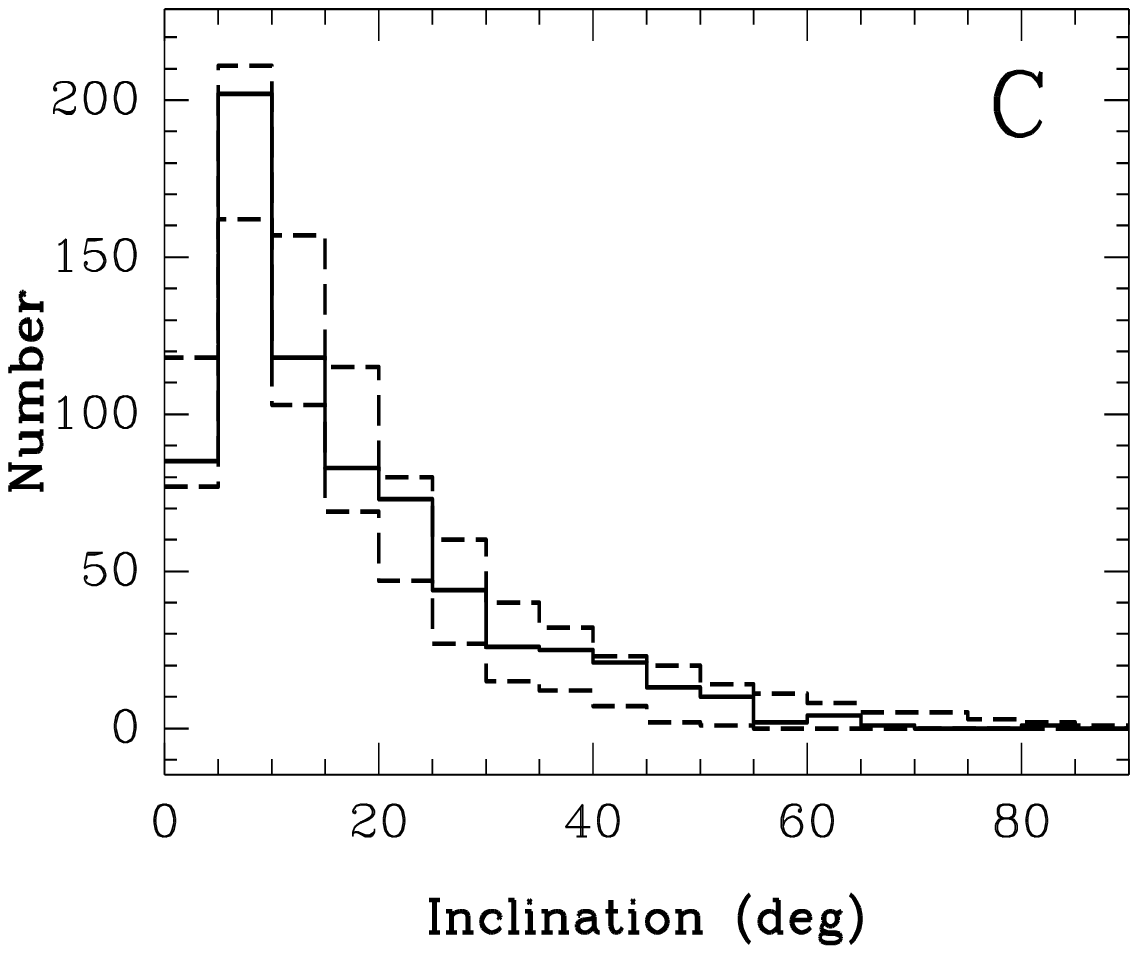, angle=0, width=4.0in, 
bbllx=120pt, bblly=200pt, bburx=480pt, bbury=510pt}
\epsfig{file=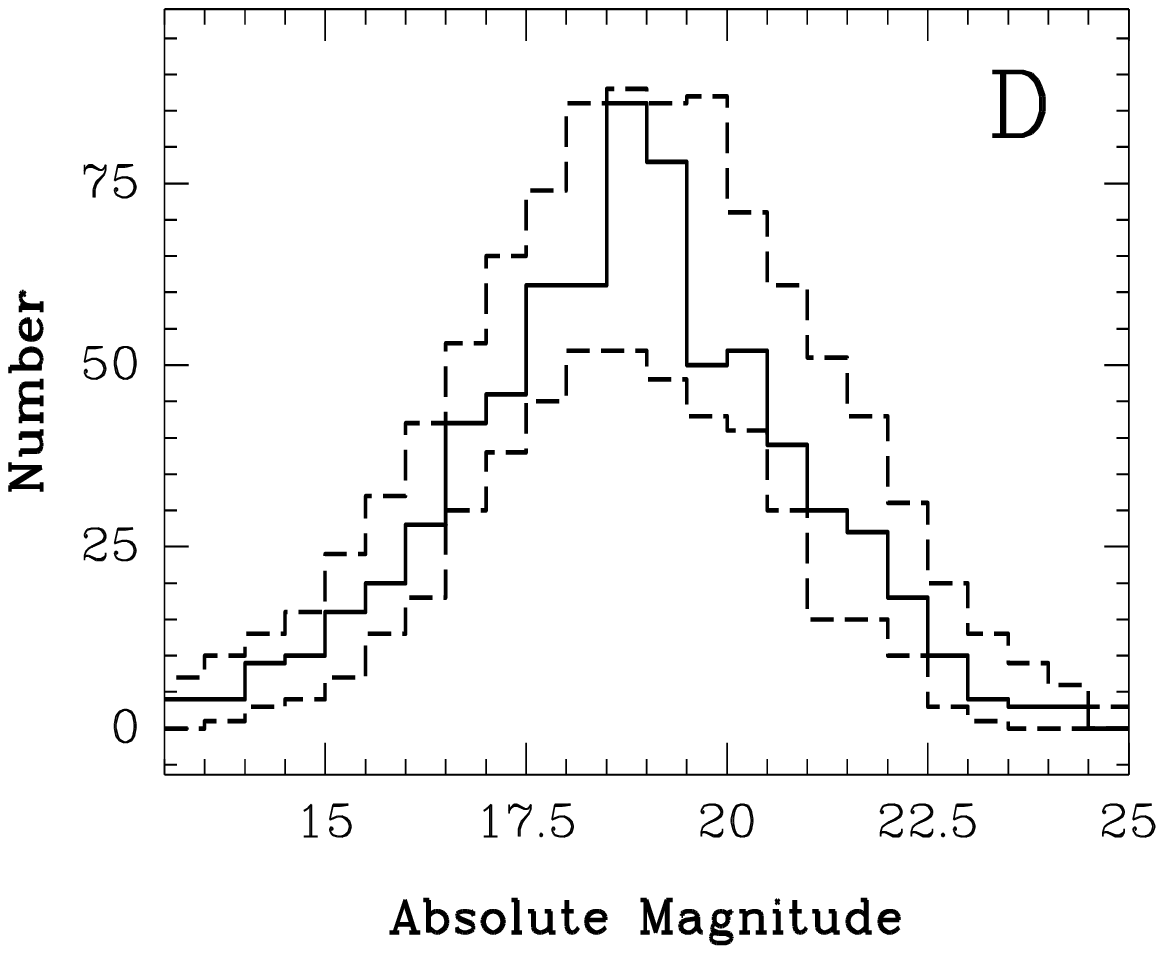, angle=0, width=4.0in,
bbllx=140pt, bblly=200pt, bburx=500pt, bbury=510pt} \\ 
\caption{Minimum and maximum (lower and upper dashed lines respectively)
distribution of 100 synthetic observed NEO populations designed to model
the distribution of NEOs known at the time of Drummond's (2000) study (solid
line). 
A) semi-major axis 
B) eccentricity 
C) inclination 
D) absolute magnitude.}
\label{fig.MinMaxNEOModels} 
\end{figure}

\placefigure{fig.MinMaxNEOModels}

To test the statistical significance of the NEO groups that we
identified (see Fig.~\ref{fig.Dru.vs.us}), we generated 100 separate
synthetic NEO models according to the $(a,e,i,H)$ distribution of
\citet{bot02a}.  Each of the three angular elements of the orbits were
generated randomly in the range $[0,2\pi)$.  We then passed each of
them through a survey simulation similar to those of \citet{jed03}
until the simulated survey detected 708 NEOs, identical to the size of
the NEO sample used in Drummond's (2000) analysis.  Very briefly, the
simulator attempts to mimic the average surveying capability of all
surveys that have identified NEOs.  It covers a 180\arcdeg\ wide range
in RA and from -30\arcdeg\ to +80\arcdeg\ in declination in the space of
14 days centered on new moon each lunation.  We found that using a
limiting magnitude of $V=20.5$ (50\% efficiency) where the efficiency
drops from 100\% to 0\% in the range from $V=19.75$ to $V=21.25$, and
a minimum rate of motion to distinguish NEOs of $0.3\arcdeg$/day,
yielded results that qualitatively agree with the observed
distributions.  The minimum and maximum range of the distributions of
the 100 synthetic populations
is shown in Fig.~\ref{fig.MinMaxNEOModels}, in which it is clear that
our simulation nicely bounds the observed distribution of NEO orbital
elements and absolute magnitudes as utilized by \citet{dru00}.

\begin{figure} 
\centerline{\epsfig{file=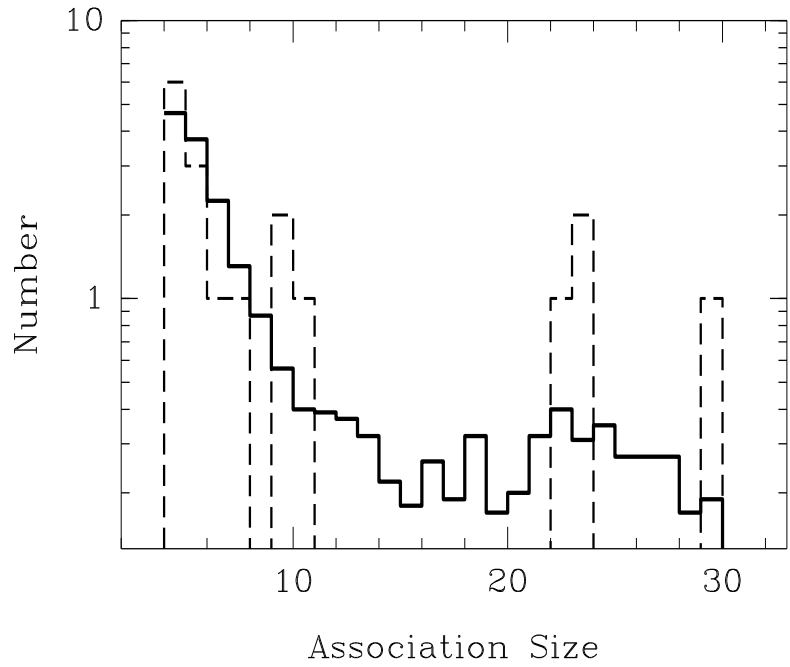,
angle=0, width=4.5in, bbllx=50pt, bblly=200pt, bburx=410pt, bbury=510pt} \\}
\caption{Average size distribution of associations detected from 100
synthetic NEO populations (solid line) and the distribution of
associations found in this analysis in the same sample as used in \citet{dru00} (dashed
line). }\label{fig.AssociationSFD} 
\end{figure}

\placefigure{fig.AssociationSFD}

We used the volume search method to identify associations in our 100
synthetic NEO populations.  In Fig.~\ref{fig.AssociationSFD} we show
the size distribution of all the associations detected in our
synthetic NEO populations and the distribution of associations found
(by our code) in the same NEO sample as used in \citet{dru00}. Since
no genetic families exist in the artificially generated NEO
populations, the fact that we identified associations containing more
than 20 members in that data, and that the size distribution of the
groupings for the real data is similar to that for the synthetic data
sets, brings into question the genetic nature of the associations
reported by \citet{dru00}.

Interestingly, both the synthetic and real data show a bi-modal
distribution. When the size of the association is $\lesssim$15 there
is a power-law drop in the SFD. For
associations containing $\gtrsim$15 members the distribution is almost
flat or modestly peaked near 22. We are unable to explain why the
technique produces this bi-modal distribution.  It is {\it not} the
case that the synthetic model may contain synthetic genetic families
{\it because} the actual NEO population from which it was derived
contains genetic families.  This is a result of the
technique used to generate the NEO model as detailed in
\citet{bot02a}.  Briefly, the correlation between the orbital elements
of members in the actual NEO population was lost in the fitting
procedure employed by \citet{bot02a}.  When we generated our synthetic
NEO population it therefore has no memory of any possible groupings in
the real NEO population.

\begin{figure} 
\centerline{\epsfig{file=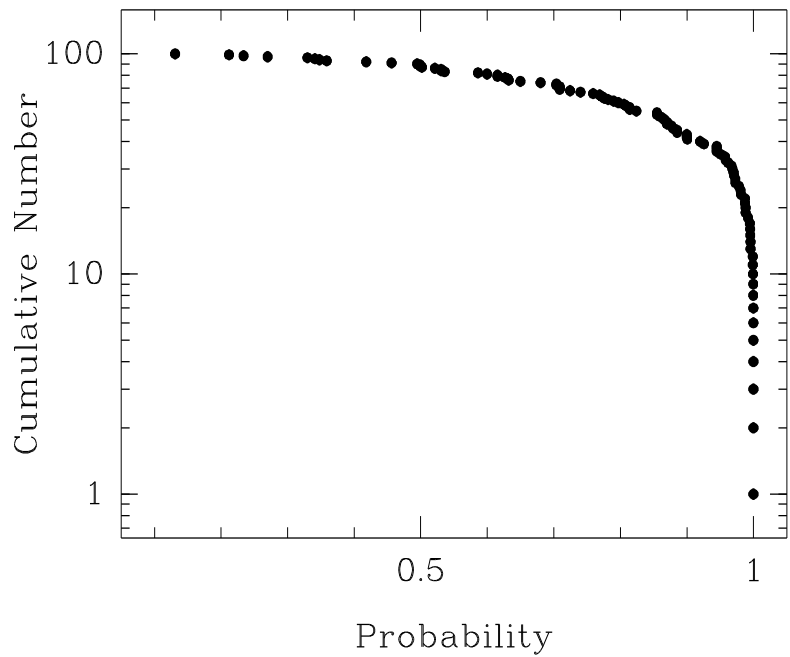, angle=0,
width=4.5in, bbllx=50pt, bblly=200pt, bburx=410pt, bbury=510pt} \\}
\caption{Cumulative distribution of the Kolmogorov-Smirnov probability that the size
distribution of associations found in a synthetic NEO population and
the one from the actual NEO population are consistent. For example, at a confidence level $\gtrsim$90\%, the size distributions of associations detected in $\sim$40 of the 100 synthetic NEO populations are statistically consistent with the same distribution found in the actual NEO population.}
\label{fig.KSTest} 
\end{figure}

\placefigure{fig.KSTest}

We test whether the real NEO population is more clustered than
the synthetic ones by determining if the two populations are different or
drawn from the same underlying distribution function.  
The Kolmogorov-Smirnov (K-S) test \citep[\eg,][]{von64,kol33,smi33}
is one of the most generally accepted techniques for testing the
consistency of two one-dimensional distributions. 
 It is applicable to non-gaussian
distributions and works well on small sample sizes. We applied the K-S test
to compare the size distribution of the associations found in the real
NEO population and that found in each
of the 100 synthetic NEO populations. We found that the size
distributions of the associations detected in 35 (54) out of the 100 synthetic
populations are consistent with that of the associations detected in the real NEO population at
a significance level $\geqslant$ 95\% (85\%).  Figure \ref{fig.KSTest}
shows the cumulative distribution of the K-S probability that
the synthetic data sets are consistent with the actual data.
Thus, there appears to be little reason to believe that any of the
associations detected by \citet{dru00} in the actual NEO sample are real.

\section{Decoherence of a Synthetic A1-like family}
\label{sec.DecoherenceOfSyntheticA1Family}

In this section we determine the length of time that a synthetic
genetic NEO family like Drummond's (2000) A1 association would be
detectable using the \dc technique.  This is important for
understanding whether the A1 association is real and also for
calculating the expected number of NEO families that may be detectable
in the future.

A significant fraction of the A1 association's mean orbit overlaps the
MB so we assume that, if it is real, it was most likely created in a
collision between a NEO and a MB asteroid 
as opposed to one of the other four
possible mechanisms identified in \S\ref{sec.intro}.  Current
computing techniques allow a realistic simulation of the entire
process of the formation and evolution of asteroid families within a
reasonable amount of clock time \citep[\eg,][]{mic02, dur04}. We have
performed a smooth-particle hydrodynamics (SPH) simulation \citep[as
in][]{dur04} of a NEO-MB asteroid collision from which an A1-like
family was formed, tracked the re-accumulation process, and then
followed the dynamical evolution of the individual fragments.  It
should be noted that some recent works \citep[\eg,][]{del04,cel04}
suggest that detailed SPH models may underestimate the ejection
velocities amongst the collision fragments.



First, we specify the parent body for the collision: assuming that the
A1 association is a genetic family we set the orbit of the synthetic
A1 parent body to the mean orbit of the actual 25 A1 members, \ie,
$(a,e,i,\Omega,\omega)=(2.21$ AU$, 0.49,
4.1^{\circ},194.3^{\circ},123.3^{\circ})$.

The lower limit on the size of the parent body is the sum of the
volume of the 25 A1 members which is dominated by the largest
fragment, 1627 Ivar.  \citet{vee89} reported radiometric observations
of this object and 2368 Beltrovata (the third largest member of A1)
and \citet{del03} provide independent radiometric observations of 1627
Ivar. Depending on the thermal model used, \citet{vee89} determined a
visual geometric albedo of 0.08 or 0.12 to 1627 Ivar and 0.05 or 0.13
to 2368 Beltrovata.  \citet{del03} found a visual geometric albedo for
1627 Ivar of 0.20, 0.15, or 0.050 depending on whether the STM, NEATM,
or FRM thermal model was used.  On the other hand, the effective
diameter for 1627 Ivar was estimated to be $8.5\pm3$ km from radar
measurements \citep{ost90}.  Given the widely varying albedo estimates
from radiometric observations and thermal modelling, we decided to use
the radar measurements as the basis for our constraints on, and best
estimate for, the diameter of the A1 parent body.

Table~\ref{tab.A1Association} shows the absolute magnitude ($H$),
taxonomic type (where available), and effective diameter (lower bound,
best estimate, upper bound) for each object in A1. The $H$ values for
each object are from the Lowell Observatory's ASTORB database
(ftp://ftp.lowell.edu/pub/elgb/astorb.html). The lower limit for the
A1 parent body is shown for three different albedo values 
at the bottom of the
table.  Note that the albedo values ($p_V = 0.07 \rightarrow 0.306$)
in this table encompass all of the albedo estimates from radiometric
measurements ($0.07 \rightarrow 0.26$) for 1627 Ivar, hence we believe
$9.8\pm3.5$ km is a suitable estimate for the size of the A1
precursor.  For the purpose of our simulations we round this number to
a synthetic A1 precursor diameter of 10 km.

We note that \citet{tan99} claim that a better preliminary estimate of
the parent body size based on geometrical considerations is obtained
merely by summing the diameters of the first and third largest
asteroids in the family.  They state that this is a much better
estimate of the parent body diameter when the volume method (as used
in this analysis and described above) yields a diameter less than the
sum of the two largest family members.  Using their technique our best
estimate for the diameter of the A1 association parent body would be
about 12 km.

With the target asteroid specified we then determine the relevant
properties of the impactor.  We assumed that the orbital elements of
the impactor are equal to those of a MB object that is likely to
collide with the A1 precursor. We identified these orbital elements in
the following manner.  The MB asteroids are almost complete to $H \sim
14.5$, so we made a histogram of the semi-major axis of all MB objects
with $H < 14.5$ that could hit the A1 precursor.  Since the A1
precursor's heliocentric distance ranges from perihelion at $\sim 1.1$
AU to aphelion at $\sim 3.3$ AU, we examined all MB semi-major axes
with 1.1 AU $< a(1-e) <$ 3.3 AU and found that the most probable
semi-major axis for the projectile is $\sim$ 3.1 AU. Similarly, the
most probable eccentricity is $\sim$ 0.11 and inclination is $\sim
10.0^{\circ}$. Therefore, we select as our impactor an object with
$(a,e,i)=(3.1$ AU$,0.11,10.0^{\circ})$.

We assume that the collision takes place at a speed equal to the 
most probable collision speed for an object with A1's
mean orbit --- 8.36 km/s (Bottke 2003, personal communication). 

\begin{figure} 
\centerline{\epsfig{file=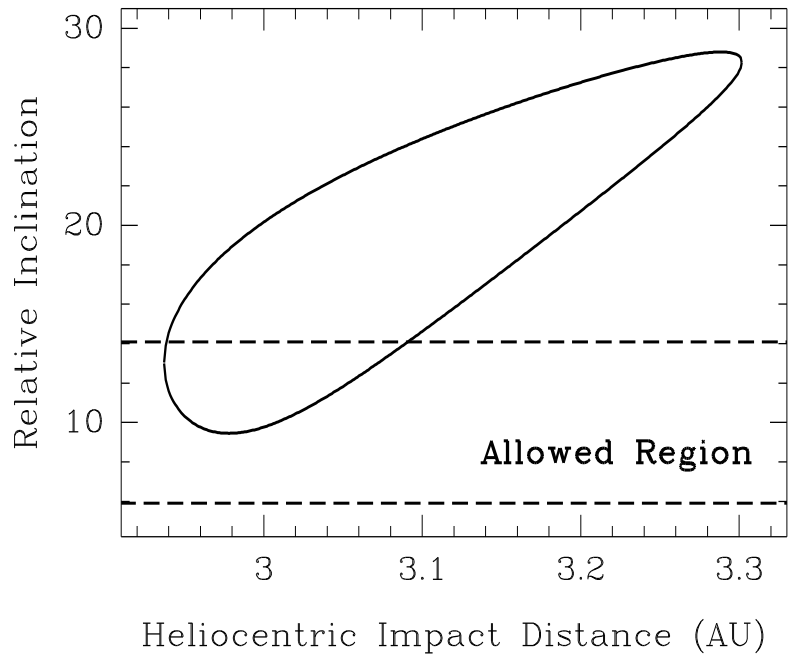, angle=0,
width=4.5in, bbllx=50pt, bblly=200pt, bburx=410pt, bbury=510pt} \\}
\caption{Relative inclination {\it vs.} heliocentric impact distance for the
  A1 precursor and projectile (see text for details). The
loop shows the relative inclinations required in order to provide the relative
impact speed (8.36 km/s) as a function of the heliocentric impact distance. The
area bracketed by the dashed lines indicates the region of possible relative
inclinations between the A1 precursor and the projectile, \ie, smaller
than the sum of and bigger than the difference between their orbit
inclinations.}
\label{fig.relativei} 
\end{figure}

\placefigure{fig.relativei}

The most likely location of an impact event is near the aphelion of the A1
precursor since it spends most of its time near that heliocentric
distance. The range of overlapping heliocentric distances between the
orbits of the A1 precursor and the projectile is from 2.94 AU to
3.30 AU, but the actual range where the impact could happen is more restricted
because we have fixed the impact speed. At higher heliocentric
distances ($> 3.09$ AU) the relative impact speed dictates that the
relative inclination (the angle between the planes of the two orbits)
must be larger than the
highest possible inclination difference, which is the sum of the two
inclinations ($14.1^{\circ}$). The loop in Fig.~\ref{fig.relativei}
shows the relationship between the relative inclination and the
heliocentric distance of the impact when constrained by our choice of
the impact speed. The dashed lines in the figure bound
the region of permitted relative inclinations between the A1 precursor and
the projectile, \ie, smaller than $i_1+i_2$ and larger than $\mid i_1-i_2
\mid$ where the subscripts $1$ and $2$ refer to the target and
projectile. For our simulations we used the smallest and largest allowable
heliocentric distances from Fig.~\ref{fig.relativei} (2.94 AU and 3.09
AU) and the one at which the
relative inclination is the smallest (2.98 AU) in order to investigate the
dependence of the decoherence time scale upon the impact location.

Since we require that the collision occurs at a specific heliocentric
distance and speed, the three angular orbital elements of the impactor can be determined
because the location of the A1 precursor at the time of collision is known and the
impactor has no choice but to cross the same point at the same time. 
We note that there
are two solutions --- the impactor may collide on either
hemisphere relative to the velocity vectors of the two objects. We present
the results for only one case since both give essentially identical
results. 

We modelled the initial stages of the impact with the 3-dimensional SPH
code SPH3D \citep{ben95}.  In keeping
with our pattern of favoring the most probable event, we
selected the smallest possible impacting
object  that would produce a barely catastrophic
disruption.  Based on our previous experience with impact simulations, and
scaling the values for the velocity of the impact and size of the target,
we selected an impactor of 0.25 km diameter.  We will argue later that
this is probably too {\it large} an impactor but we believe that the
results of the simulation are still of practical utility.  Similarly, the most probable impact angle is
$45^{\circ}$ as used in this simulation. 

The 10 km diameter target was comprised of 100,000 SPH particles and, to roughly match the SPH
particle volume density of the target, the impactor is comprised of 700
particles. In this simulation the A1 precursor was catastrophically
disrupted and shattered into thousands of independent fragments by the
impactor.  The SPH phase of the simulation followed the impact for 100 seconds
after which time the ejecta flow fields
were well established and no further fragmentation or damage was observed. 


The outcome of the SPH simulation was handed off as the initial conditions
for an N-body simulation using the cosmologically derived pkdgrav code
\citep{ric00, lei00, lei02} that followed the trajectories of the ejecta
fragments for sufficient time to allow gravitational re-accumulation of the
larger collision fragments.  This technique was used by \citet{dur04} to
model the formation of asteroid satellites resulting from large impacts on
100-km scale MB asteroids. With the small sized target assumed here
for the A1 precursor, there is not much gravitational re-accumulation among
the largest fragments and the N-body phase of the simulation need only be
run for 200 N-body timesteps (each step corresponding to about 50 s)
in order to establish fragment positions and ejection velocities.


From the velocity vectors of the A1 precursor and the impactor at the
collision location we can derive the three Euler angles between the impact
space, where the simulation was performed, and the external
heliocentric orbital space in which we follow the post-collision dynamical evolution
of the fragments.
The initial orbits of the re-accumulated fragments are handed over to
Mercury6 \citep{cha99} to follow their dynamical evolution. The SPH simulation produced
1243 fragments down to a diameter of 216 m (pre-determined by the resolution of
the simulation).
Note that the smallest diameter amongst known members of the real A1
association is only $130\pm50$ m (see Table~\ref{tab.A1Association}).  

Only a fraction of the fragments would be discovered in reality, so we
applied an {\it ad hoc} observational selection function to determine
which of the synthetic fragments were detected.  The correction factor is a
complicated function of a survey's performance characteristics and the
orbital element distribution of the detected objects \citep{jed02}.  Since
all the synthetic A1 family members have essentially identical orbital
elements we would like to determine the selection effect as a function of
$H$ for all the surveys that contributed to finding NEOs with orbital
elements similar to the A1 mean orbit.  This would be difficult if
not impossible to determine in practice.  Instead, we chose simply to 
divide the set
of all known 708 NEOs at the time of Drummond's (2000) analysis as a
function of absolute magnitude, $n(H)$, by the debiased ``true"
distribution of NEOs according to \citet{bot02a}.  We fit the resulting
distribution to a function of the form: 
\begin{equation} 
P(H) = \frac{1}{e^{\frac{H-L}{W}}+1} \,, 
\label{eq:selection} 
\end{equation} 
and found $L= 15.82\pm0.19$, $W = 1.15\pm0.11$. $P(H)$ gives the probability
that an NEO with absolute magnitude $H$ would have been discovered by the
time of Drummond's (2000) analysis. 

\begin{figure} 
\centerline{
\epsfig{file=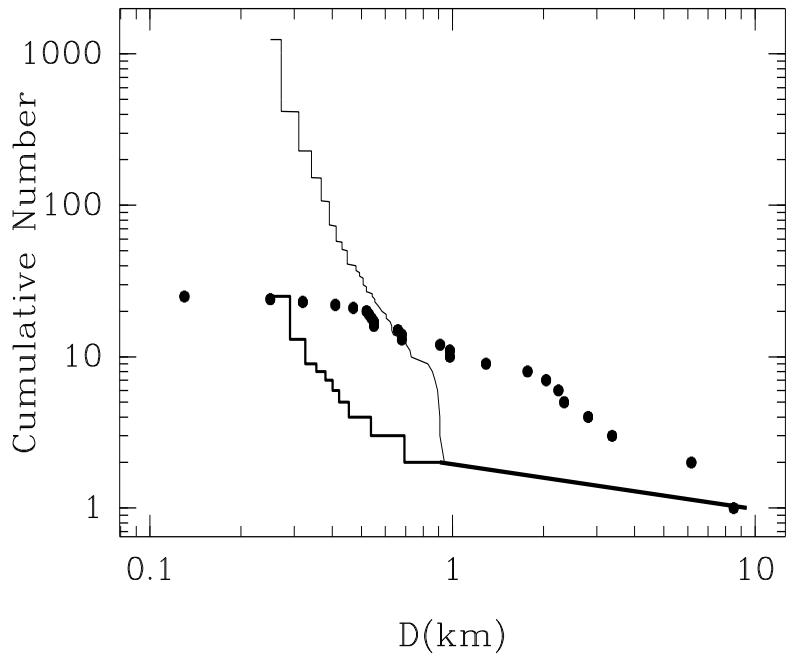, angle=0,
width=4.5in, bbllx=50pt, bblly=200pt, bburx=410pt, bbury=510pt} \\}
\caption{Cumulative size distribution of the actual A1 association
(dots) and the synthetic A1-like family before (thin line) 
and after (thick line) applying an {\it ad hoc} observational selection effect.}
\label{fig.SyntheticA1.vs.A1} 
\end{figure}

\placefigure{fig.SyntheticA1.vs.A1}

We applied this discovery probability with the central values for $L$
and $W$ to the re-accumulated synthetic A1 family.  An albedo of 0.128
(the same as was used for the best estimate of the size of the A1
precursor as in Table~\ref{tab.A1Association}) was used to convert the
size of the objects into an absolute magnitude.  Running many trials
for this process (which objects ``survive" the selection function of
equation \ref{eq:selection} is random) yielded a mean size of an
observed synthetic A1 family of $25\pm4$ members compared to the 25
members of the actual A1 association reported by \citet{dru00}.  The
size distribution of our synthetic population is compared to the
actual size distribution in Fig.~\ref{fig.SyntheticA1.vs.A1}.

It appears that our model does not reproduce the actual A1 association's
size distribution.  This could be due to our having selected an
impactor that caused far too much damage to the target, shattering it
with such violence that the fragments were unable to gravitationally
accumulate into larger objects.  However, considering that our
dynamical integrations of the fragments consider only gravitational
effects, the SFD of the fragments is probably not important to the
remainder of this study.

We now examine the decoherence of the synthetic A1 family as its
members dynamically evolve under the gravitational influence of the
Sun and eight planets (Mercury through Neptune). It was only
necessary to integrate the motion of the 25 fragments that were detected
rather than the ensemble of 1242 fragments. The orbits were propagated forward in time using a hybrid symplectic/Bulirsch-Stoer integator (Mercury6, Chambers 1999) with a step-size, $\tau$, of 30 days (\ie, 1/40 of the
shortest initial orbital period amongst the fragments). We verified the result by comparing it to an identical run on the same fragments using the general Bulirsch-Stoer integrator with the same step-size. 

In reality, this evolution occurs
within a population of background objects that are also
interacting gravitationally with the major objects in the solar system
and with each other. In order to save time, for the purpose of this study, we decided not
to integrate the motion of all the background objects.  Instead, we
used two different static background NEO populations: B1 was a
randomly selected synthetic NEO population (from
\S\ref{sec.StatisticalSignificanceOfNEOGroups}) with 708 objects and
B2 was formed using the same 708 NEOs as used in Drummond's study but
removing the 25 actual A1 members.  This means that our static background
populations possessed a slightly different mean density in orbital
element space with a total of 733=708+25 objects in the B1 case and
only 708 objects in the B2 case.  This introduces a difference in the
mean density of only a few percent.  It would be difficult to correct
because it would require selectively removing 25 members from the
synthetic B1 population that somehow match the original A1 members.
Instead, we chose to ignore this small difference in density and note
that, in any event, it makes the identification of groupings in the
synthetic B1 data slightly more likely.  We introduced our synthetic
A1 family members into the background NEO populations and search for
groupings every 2,000 years as the A1 family evolves in time. 
Since our purpose was to track the evolution of the A1 family,
we only performed the volume search
for candidate groupings around the 25 detected fragments as opposed to
searching the entire sample. The mean orbit of the resulting
association may be beyond the boundary of the original fragments
because interloping asteroids may be incorporated into the final
association due to the iterative aspect of the volume search
technique.

While our definition of associations, strings and pairs matches that
of \citet{dru00} our manner of selecting the final groups may be
different.  We searched for associations using the (known) synthetic
A1 members as seeds and select the association that includes
the most members of the original synthetic family.  If multiple
associations meet this criteria then we selected the association
containing the smallest number of interlopers.  Similarly, strings
were identified using the synthetic A1 members as seeds and selecting
the string that contains the most members of the original
association.  Once again, if multiple strings met this criterion we
used only the string that contained the least interlopers.  Finally,
pairs were identified only for objects within the original synthetic
A1 association in the current timestep.

\placefigure{fig.SyntheticA1Evolution}

Fig.~\ref{fig.SyntheticA1Evolution} demonstrates the decoherence of an
A1-like family from creation until it is 1 Myr old. The results for
other choices of the 25 detected fragments, the NEO background and for the heliocentric distance
of impact appear nearly identical.  It shows that the
rapid orbit evolution of NEOs mixes the genetically related synthetic A1
objects and background objects within $\sim$300 Kyr so that the association
becomes undetectable to this technique. The fact that all our runs
have similar results indicates that the
decoherence time scale is independent of the choice of the background
NEO population and the location where the precursor
is disrupted.    

\section{Decoherence of the Actual A1 Association}
\label{sec.ActualA1Evolution}

The A1 association is the largest among the 14 identified by
\citet{dru00} and all its 25 members are included in the largest
association identified in this study (see Table
\ref{tab.LargeGroupComparison}).  It is the most statistically
significant association found using our technique and the 25 common
members have a mean orbit of $(a, e, i) \sim$ (2.21 AU, 0.49,
$4.1^{\circ})$.  In \S\ref{sec.DecoherenceOfSyntheticA1Family} we
argued that the A1 association must have derived from a parent body
$\gtrsim 10$ km in diameter if it represents a genetic NEO family, yet we
calculated in \S\ref{sec.intro} that the production of a family from
the disruption of an object this large is extremely unlikely.
However, just because an event is unlikely doesn't mean that it did not
occur.

In this section we use two different methods to examine the
possibility that Drummond's (2000) A1 association is a bona fide genetic
family or due merely to chance alignments or observational biases.
First, we integrate forward the orbits of all the A1 members and see
how long it remains detectable as an association.  If the association
is real we would expect that the members remain an association for
some time into the future.  Second, we integrate the orbits {\it
backward} in time to see if their angular orbital elements converge,
indicating a common point of origin at some time in the past. The
backward integration is justified for this association because we know
that, if it is a genetic family, it can not be very old.  The latter
technique has been used by \citet{nes02} to determine the age of the
Karin cluster within the MB.


The orbits of the 25 members in A1 were propagated forward (and
backward, \S\ref{sec.A1.backwards}) in time using the general
Bulirsch-Stoer integrator with $\tau = 20$ days (\ie, 1/40 of the
shortest initial orbit period amongst the actual A1 members). The
current orbital elements of the A1 members are extracted from the MPCORB
database as of 2005 March 1 (ftp://cfa-ftp.harvard.edu/pub/MPCORB/) that contains accurate orbital elements
derived from multi-opposition observations of the NEOs.
Once again, the integration took
into account the gravitational perturbations due to all major planets (Mercury
through Neptune) but neglected the evolution of background NEOs. 
Every 2,000 years we searched for groupings (pairs, strings
\& associations) in the same sample as used by \citet{dru00} using the techniques described above but replacing the 25 A1 members with their evolved orbits.

\placefigure{fig.ActualA1Evolution}

For the same reasons as in
\S\ref{sec.DecoherenceOfSyntheticA1Family}, we only searched for density
enhancements above the local background using the 25 A1 members as
seeds in the volume search. We searched for strings and pairs within
the A1 association at each time step in a manner identical to that
described in \S\ref{sec.DecoherenceOfSyntheticA1Family}.

Fig.~\ref{fig.ActualA1Evolution} shows the evolution of A1 during the next 500
Kyr. The A1 members rapidly evolve away from each other as shown in
Fig.~\ref{fig.ActualA1Evolution}A --- the average \dc amongst all A1 members
increases from 0.2 to $>$0.6 in $\sim$20 Kyr and to nearly 1.0 after
slightly more than 200 Kyr.  At the first time step (\ie,
now), all the 25 A1 members are included in our 29-member association
(Fig.~\ref{fig.ActualA1Evolution}B), but after just two time steps
(4,000 years) the association is essentially undetectable.  In the
following 500 Kyr the largest A1-like associations contain anywhere
from a few to about 30 objects but those groups contain only a few of
the original A1 members.  In other words, those associations are quickly
and heavily contaminated by interlopers.

The situation is similarly bleak for strings (Fig.~\ref{fig.ActualA1Evolution}C) and pairs (Fig.~\ref{fig.ActualA1Evolution}D) detected within
the associations at each time step.  The largest string detected at the
first timestep contains only 5 members and
after just 20,000 years not a single time step ever contains more than 4 members of the
original A1 cast.  Similar behavior is observed for the pair counts
within the original A1 members.  With 25
objects there are 300 possible unique pairings but the maximum number
of pairs within the A1 association is 10 for only the very first time step.

The result that the A1 association is essentially undetectable within 
$\sim$10 Kyr suggests that it too is merely a statistical fluctuation in the
density of NEO orbital elements and argues against the genetic nature of
A1.  It is extremely unlikely that such a large association would be
detected just as this technique lost its ability to identify it due to
the orbital decoherence of its members.

\subsection{Evolution of the A1 association's $\Omega$ and $\omega$}
\label{sec.A1.backwards} 


\citet{nes02} integrated the orbits of members of the Karin cluster
backwards in time to identify when the longitude of ascending node
($\Omega$) and argument of perihelion ($\omega$) of all members were tightly
clustered.  These angular orbital elements evolve rapidly and
secularly under the influence of the gravitational perturbations of
the planets.  Determining the point in time at which the values were
tightly clustered allowed them to accurately date the age of that
small and new asteroid family.  The evolution of these angular
elements will be considerably faster for NEOs but we tested for their
convergence despite the low probability of success.

\placefigure{fig.dif.peri.node}

Figure \ref{fig.dif.peri.node} shows the evolution of the node and
perihelion angles for the last 50 Kyr.  We only show the evolution
over this reduced time period because it is representative of the
entire evolutionary period of this association (see next paragraph for
discussion).  Note that at $t=0$ the values are not consistent with
being randomly distributed since the \dc technique specifically
selects sets of objects with non-random orbits (in $\Omega$ and
$\omega$ as well as $a$, $e$ and $i$).  All 25 values of the
longitude of ascending node are distributed in the restricted range
$125\arcdeg \lesssim \Omega \lesssim 325\arcdeg$ and while the
longitude of perihelion is similarly restricted in the range $0\arcdeg
\lesssim \Omega \lesssim 180\arcdeg$.

\placefigure{fig.peri.node.range.stdev}

To evaluate whether these orbital elements become less random in the
past, in other words, more ``clumped", we determined the probability
that the elements were consistent with being flatly distributed using
the K-S test at each time step.  This test was uninformative, the
distribution
changing rapidly and randomly from consistent to inconsistent over the
entire 300 Kyr period.  The problem is that there are many ways for a
distribution to be ``clumped".  Instead, we resorted to examining the
distribution of $\Omega$ and $\omega$ in terms of their range, and RMS
spread within that range, as a function of time as shown in Fig.
\ref{fig.peri.node.range.stdev}.  The range of the two elements and
their
RMS spread within that range is at a minimum at the present time.  In
other words, at no time in the past were these
angular elements more ``clumped" than at the present time.  The
``clumpiness" at the present time is easily explained as a consequence
of the \dc technique selecting clumps in their orbit distribution.  Once again,
this result contradicts the hypothesis that the A1 association
represents a genetically linked set of asteroids.

\section{Identifying Actual Genetic Families}
\label{sec.IdentifyingActualGeneticFamilies}

The fact that the volume search method identified a number of
``associations" in synthetic NEO populations where no genetic
families exist (\S\ref{sec.StatisticalSignificanceOfNEOGroups})
indicates that the method, by itself, can not distinguish between
genetic families and random, over-dense regions in the orbit element
space. In this section we show that, in principle, it should be possible
to separate the real genetic families from the background by combining
results from the volume search and string/pair searches.

\placefigure{fig.IdentifyingActualGeneticFamilies}

\begin{figure}
\epsfig{file=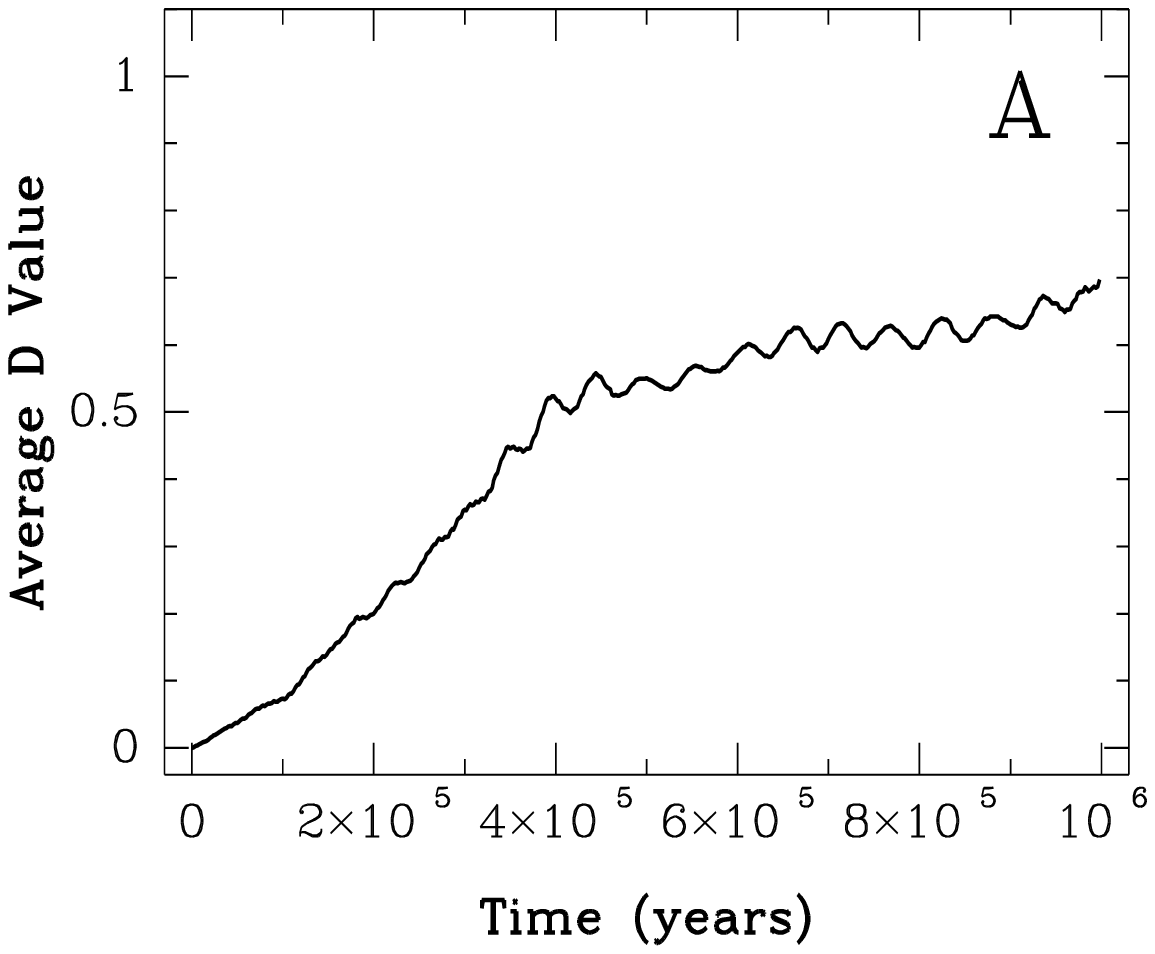, angle=0, width=4.0in,
bbllx=120pt, bblly=200pt, bburx=480pt, bbury=510pt}
\epsfig{file=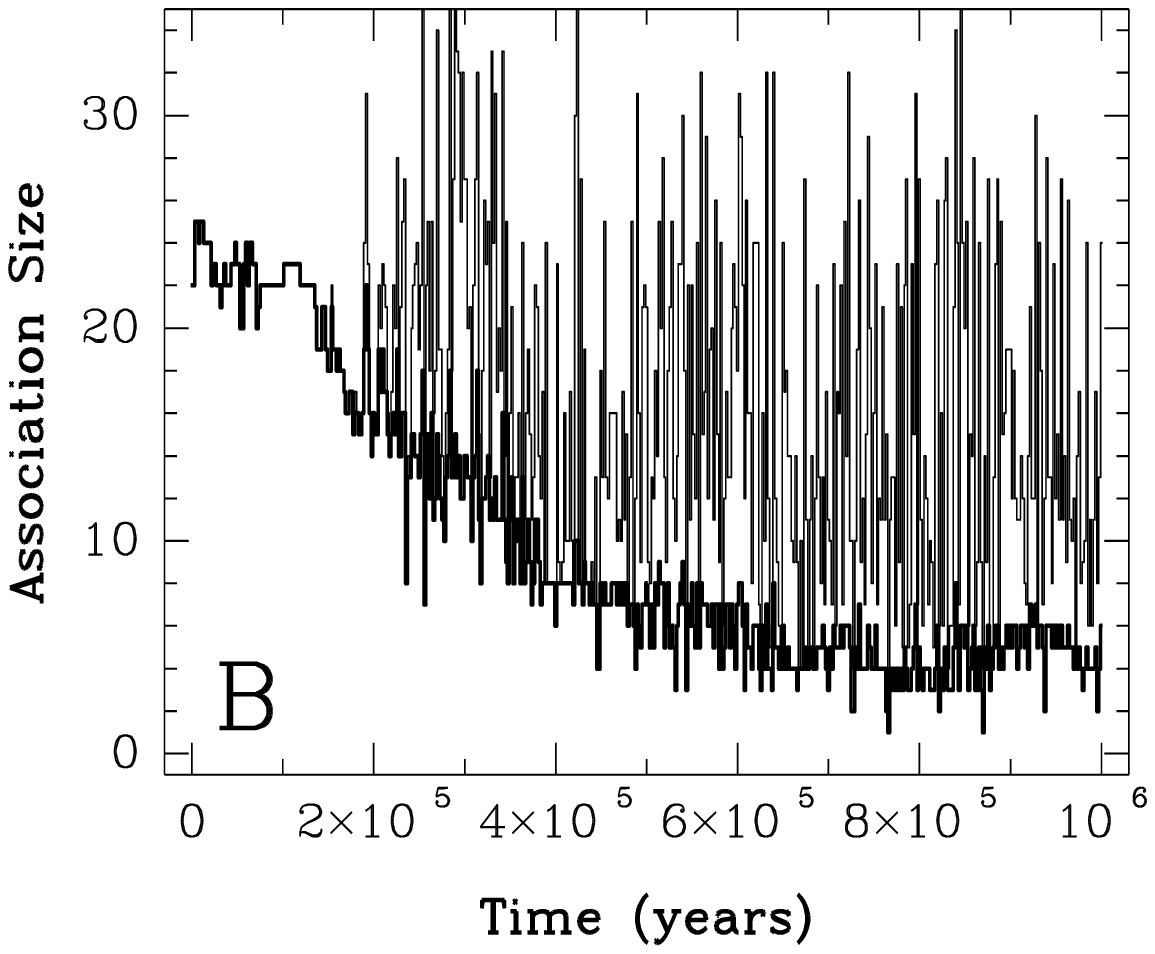, angle=0, width=4.0in, 
bbllx=140pt,bblly=200pt, bburx=500pt, bbury=510pt} \\ 
\epsfig{file=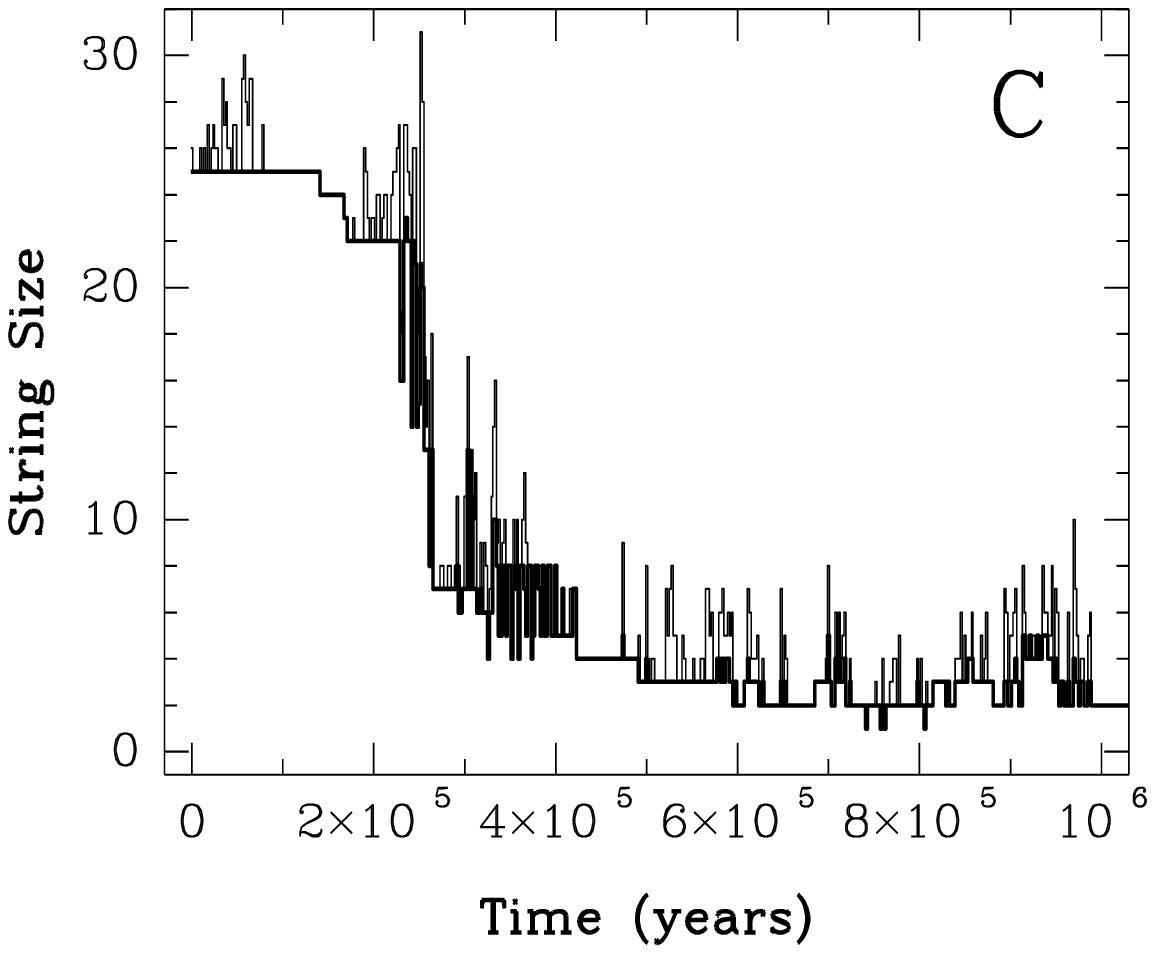, angle=0, width=4.0in, 
bbllx=120pt, bblly=200pt, bburx=480pt, bbury=510pt}
\epsfig{file=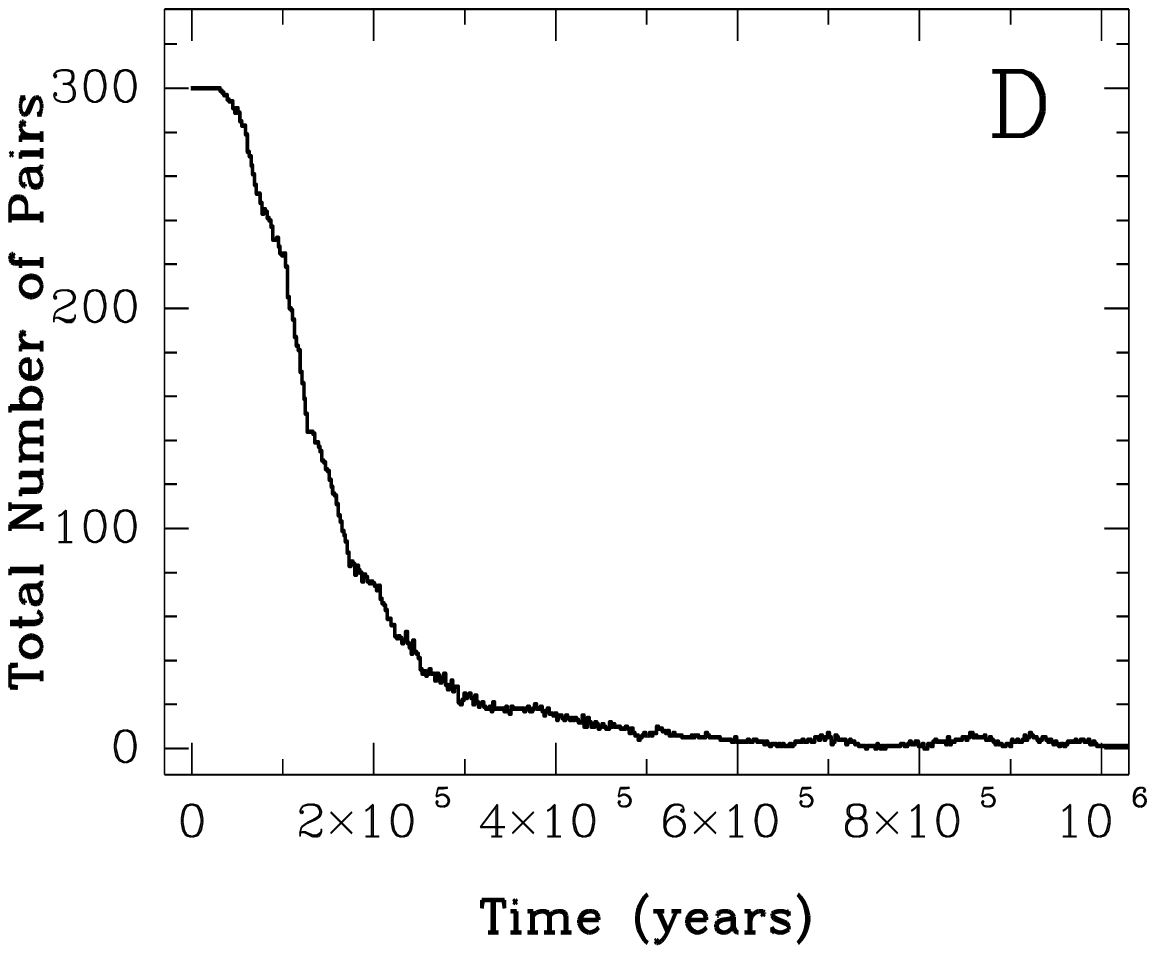, angle=0, width=4.0in,
bbllx=140pt, bblly=200pt, bburx=500pt, bbury=510pt} \\ 
\caption{Evolution of a synthetic A1-like association in the next 1 Myr.  These results use the B1 background population and an impact heliocentric distance of 2.98 AU for the collision (see text for details).
A) the mean \dc value among all synthetic A1 members, 
B) the largest detectable A1-like
{\it association} (thin line) and the number of objects that are original
members of the synthetic family (thick line) in each detected group, 
C) the largest {\it string} with $D_{string} = 0.1$ (thin line)
and the number of objects that are original members of A1 (thick line), 
D) total number of all {\it pairs} within the
original synthetic A1 members meeting the $D_{pair} = 0.1$ threshold.
}\label{fig.SyntheticA1Evolution}
\end{figure}

\begin{figure}
\epsfig{file=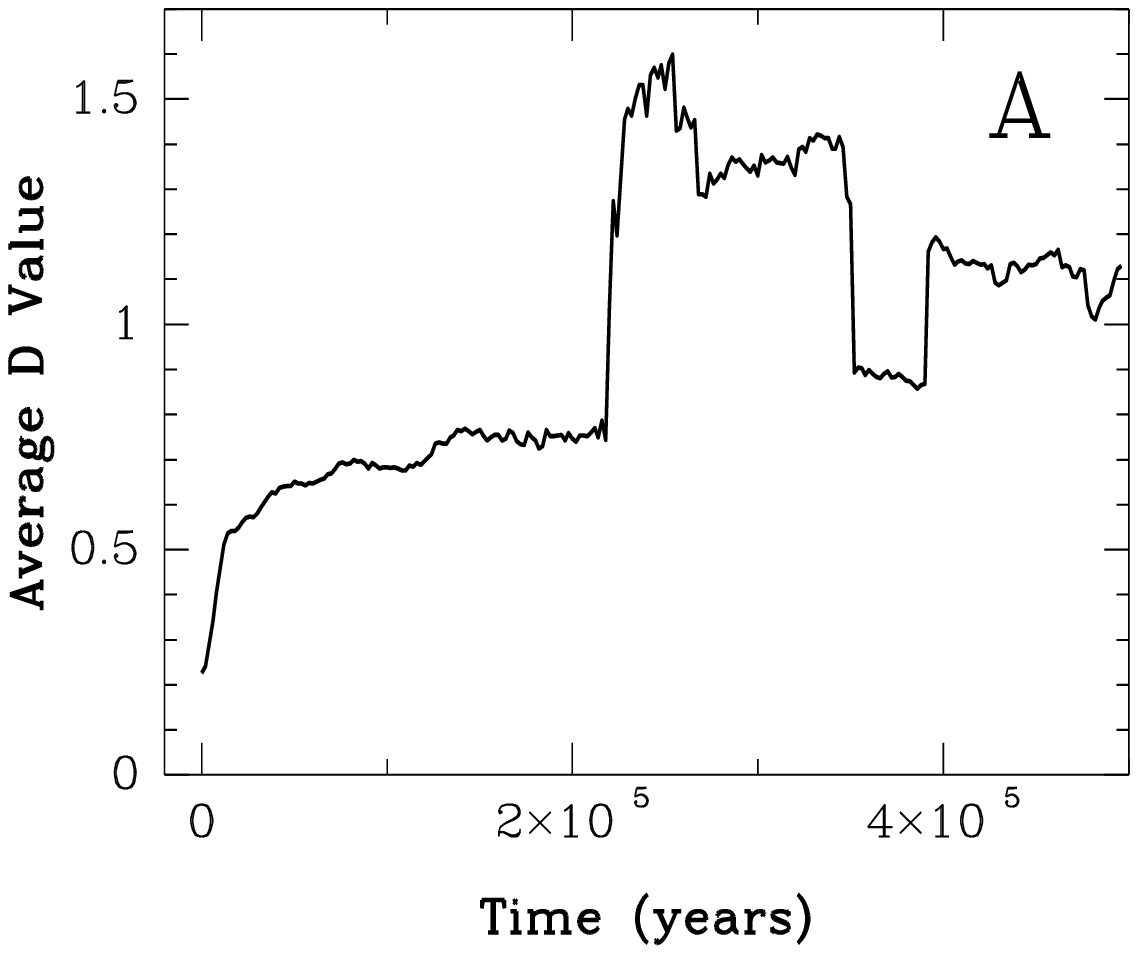, angle=0, width=4.0in,
bbllx=120pt, bblly=200pt, bburx=480pt, bbury=510pt}
\epsfig{file=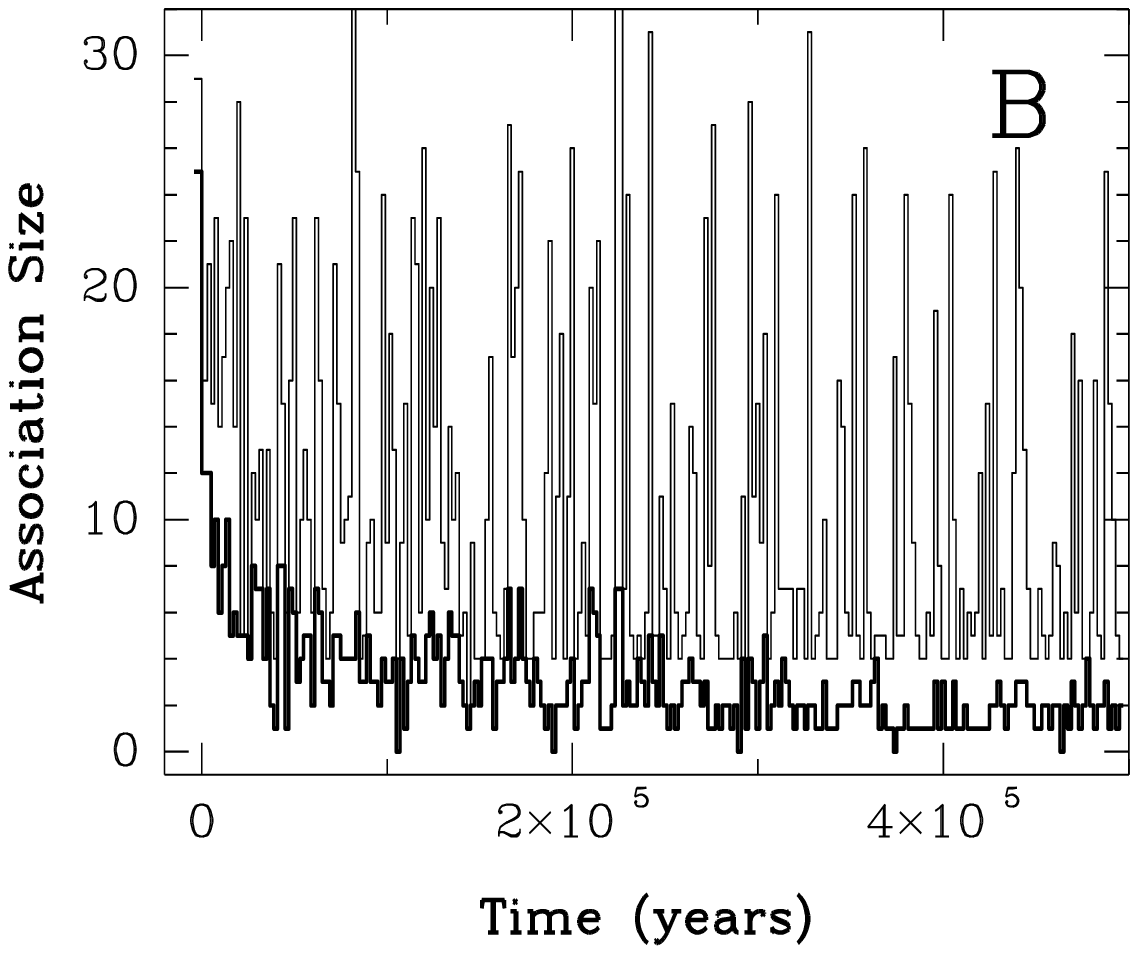, angle=0, width=4.0in, 
bbllx=140pt,bblly=200pt, bburx=500pt, bbury=510pt} \\ 
\epsfig{file=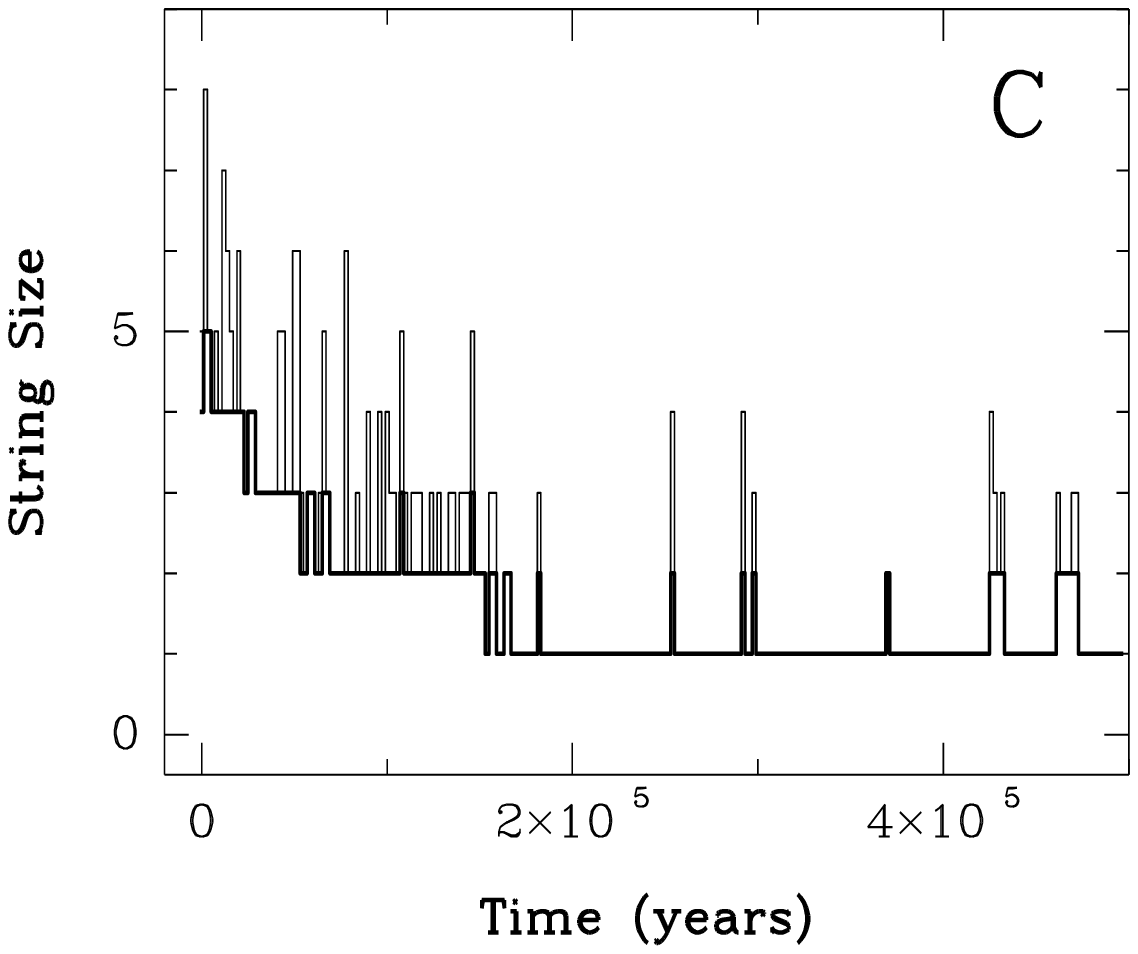, angle=0, width=4.0in, 
bbllx=120pt, bblly=200pt, bburx=480pt, bbury=510pt}
\epsfig{file=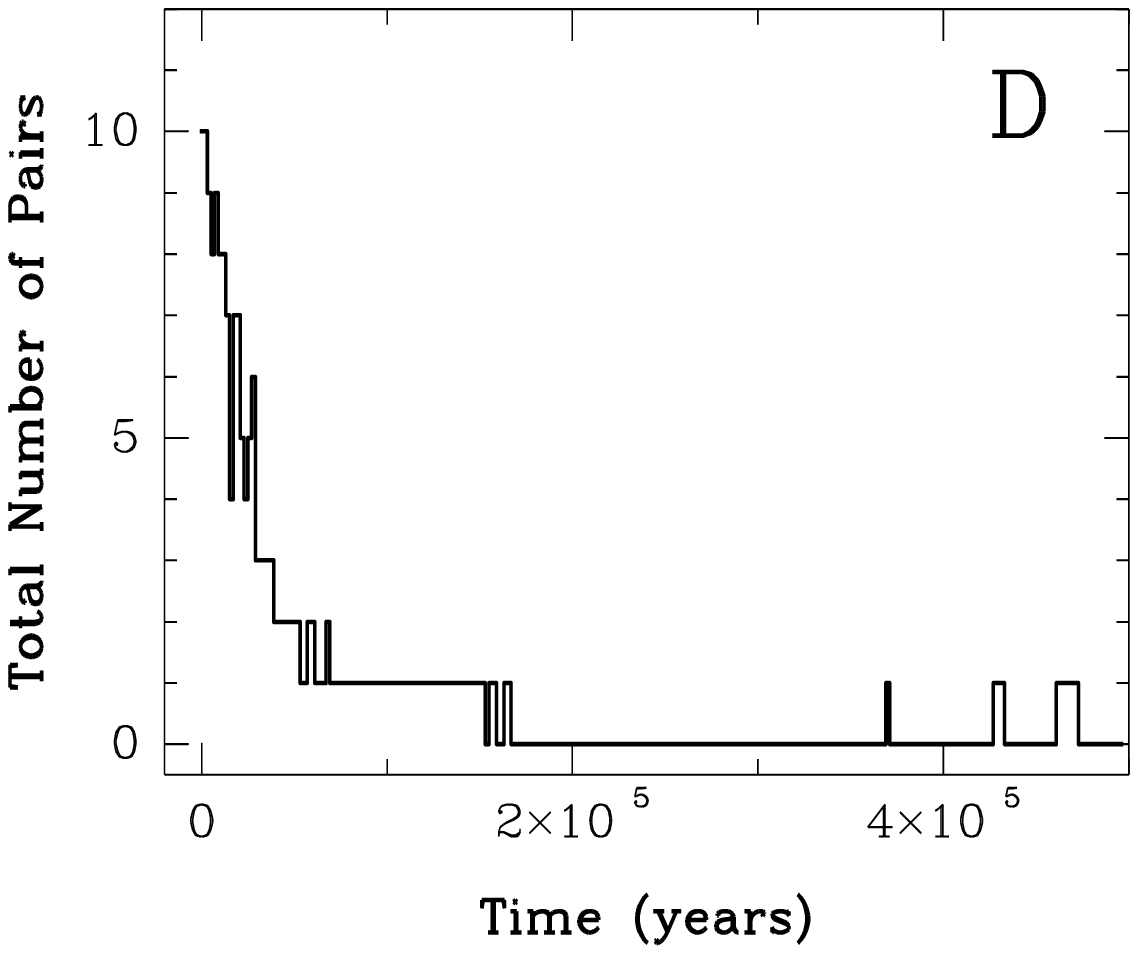, angle=0, width=4.0in,
bbllx=140pt, bblly=200pt, bburx=500pt, bbury=510pt} \\  
\caption{Evolution of Drummond's (2000) A1 association in the next 500
  Kyr ---  The sub-figures are otherwise identical in description to
  those in Fig.~\ref{fig.SyntheticA1Evolution} but are reproduced here for clarity.
A) the mean \dc value among all original A1 members, 
B) the largest detectable A1-like {\it association} (thin line) and the 
number of objects that are original A1 members
detected in each detected group (thick line), 
C) the largest {\it string} with $D_{string} = 0.1$ (thin line) 
and the number of objects that are members of the original A1
  association (thick line), 
D) total number of all {\it pairs} amongst the
original A1 members meeting the $D_{pair} = 0.1$ threshold.  There
are a total of 300 ($C_{25}^{2} = 25 \times 24 / 2$) possible pairs within the
25-member A1 association.}
\label{fig.ActualA1Evolution} 
\end{figure}

\begin{figure}
\epsscale{.90}
\plotone{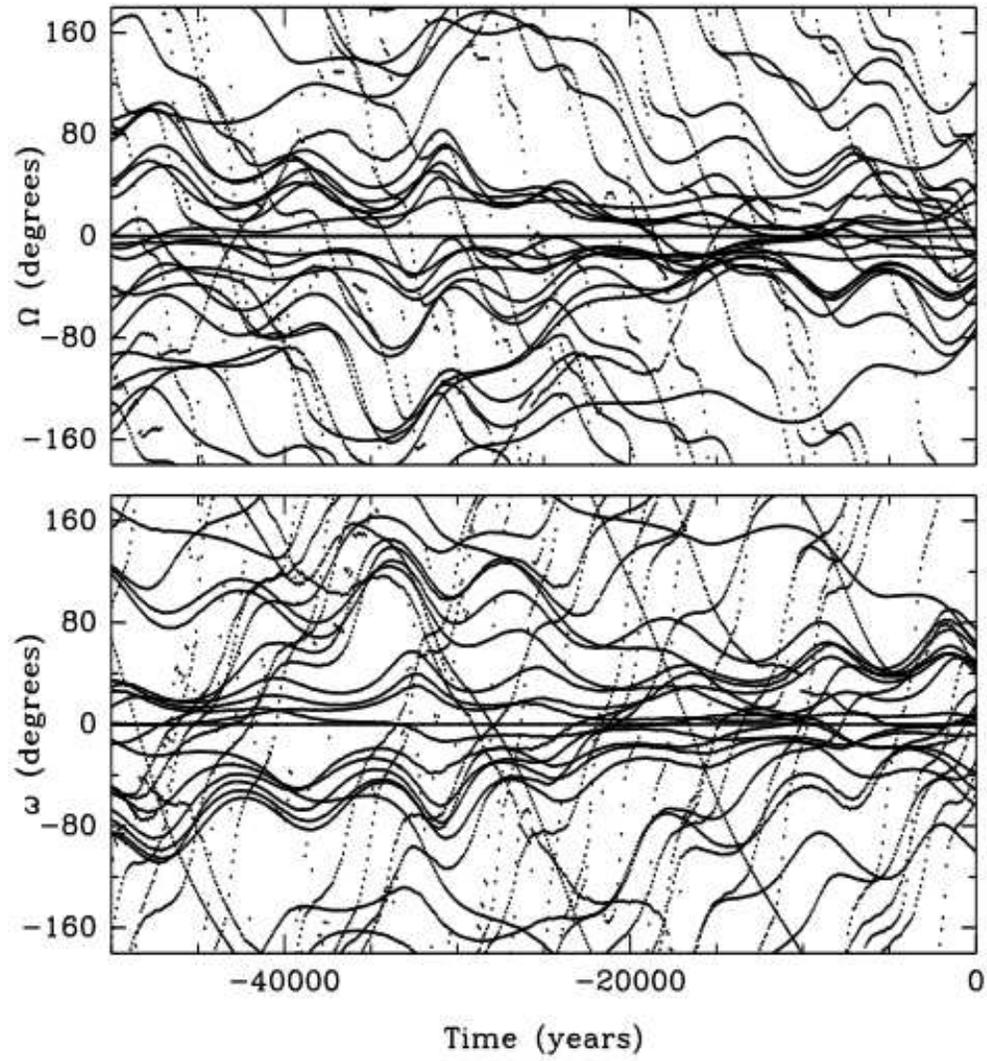}
\caption{Time evolution of the longitude of the ascending node,
  $\Omega$ (top), and of the argument of perihelion, $\omega$ (bottom), for
  the 25 members of Drummond's (2000) A1 association.  The orbital
  elements are shown relative to the value for 1998 QQ52 (that appears
  as the straight line at 0$\arcdeg$) because its orbit is the
  closest to A1's mean orbit.}
\label{fig.dif.peri.node}
\end{figure}

\begin{figure}
\epsfig{file=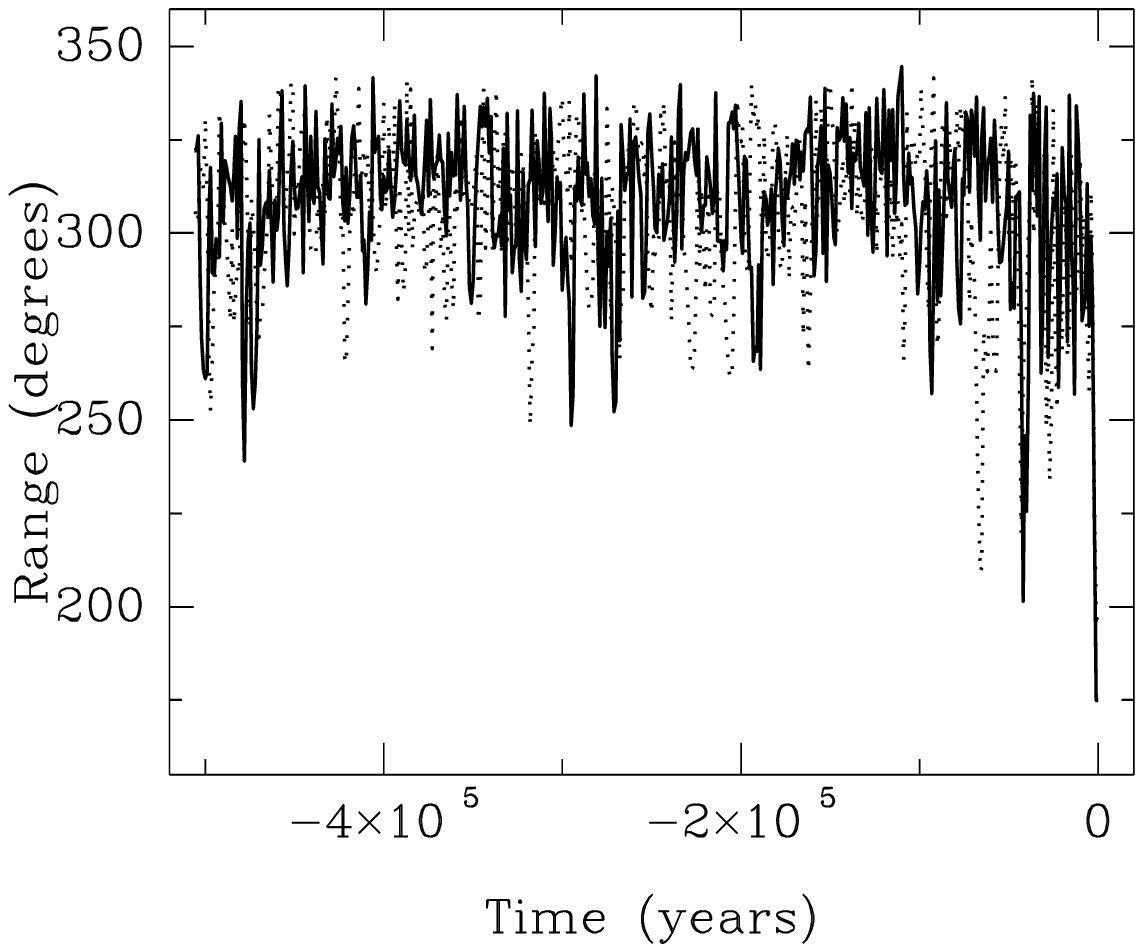, angle=0, width=4.0in,
bbllx=120pt, bblly=200pt, bburx=480pt, bbury=510pt}
\epsfig{file=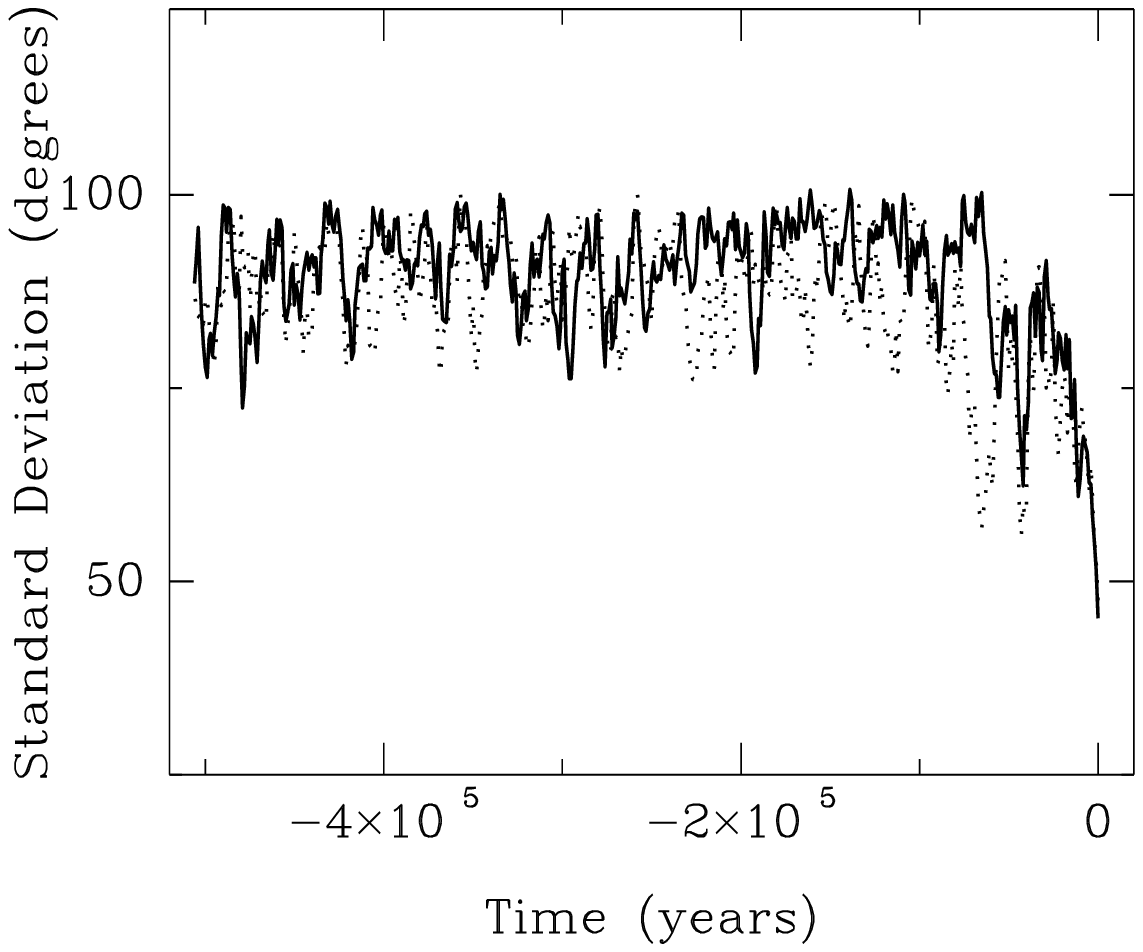, angle=0, width=4.0in, 
bbllx=140pt,bblly=200pt, bburx=500pt, bbury=510pt} \\
\caption{{\it Left} -- evolution of the range (max -- min) of $\Omega$ (dotted) and $\omega$ (solid) in the last 500 Kyr for Drummond's (2000) A1
  association.  {\it Right} -- evolution of the RMS spread of the same
  angular elements.  In creating these figures we
  have taken into account the wrap-around at the 360\arcdeg\ $\rightarrow$ boundary.}
\label{fig.peri.node.range.stdev}
\end{figure}

\begin{figure}
\epsfig{file=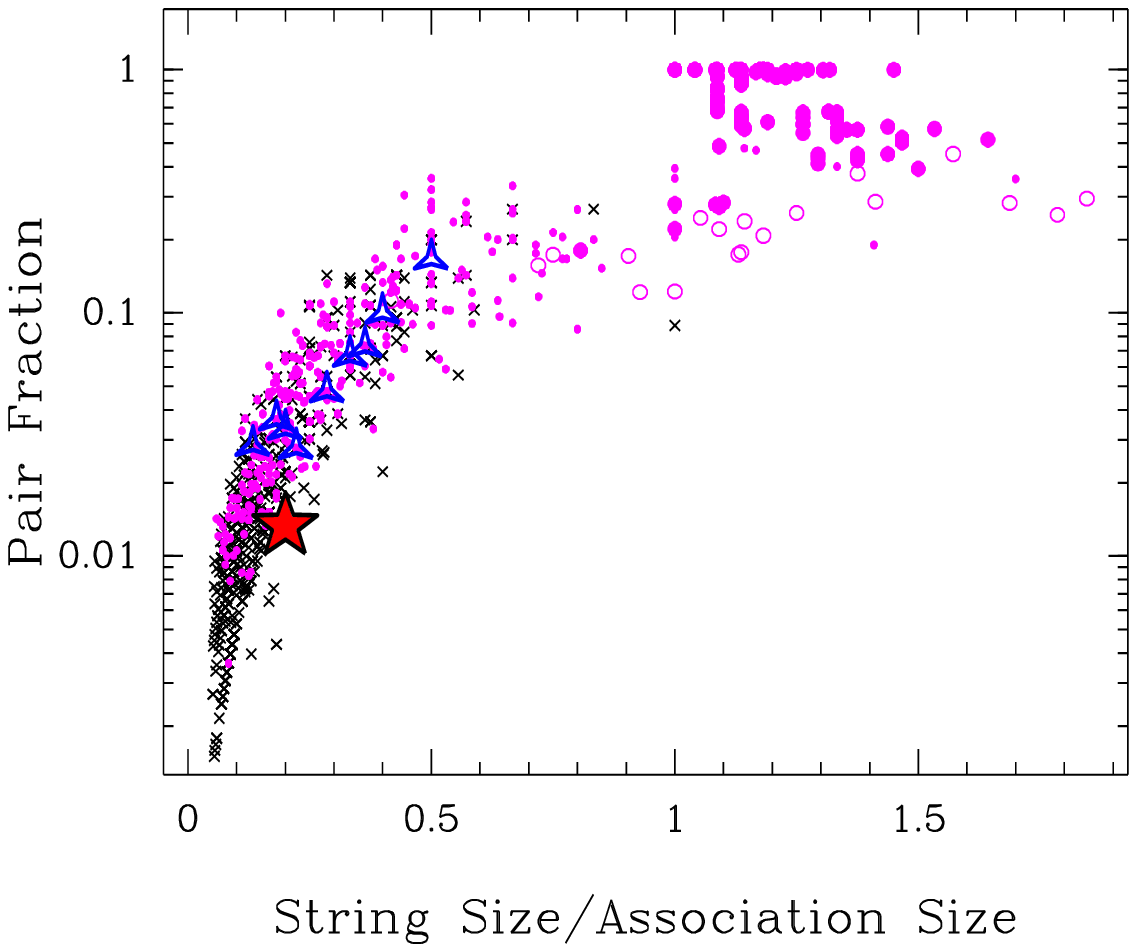, angle=0, width=4.0in,
bbllx=120pt, bblly=200pt, bburx=480pt, bbury=510pt}
\epsfig{file=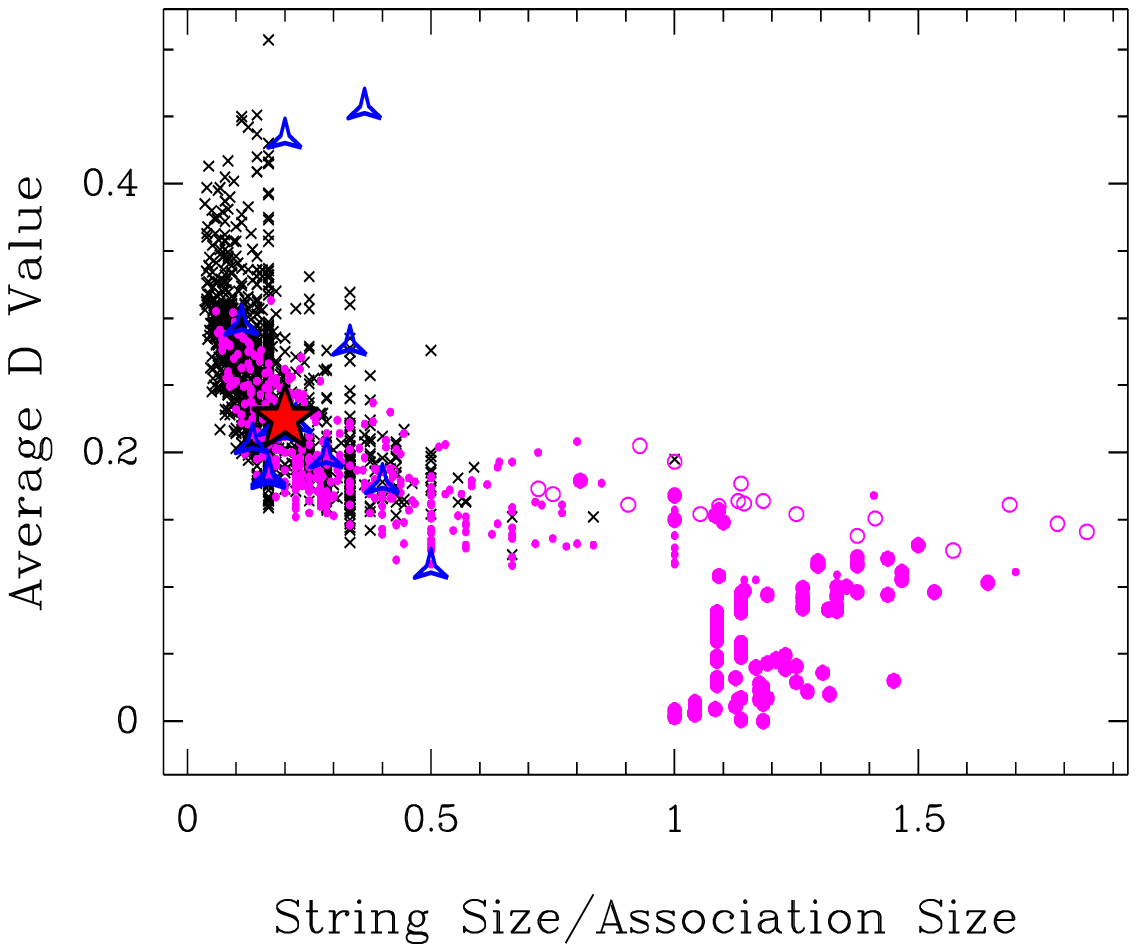, angle=0, width=4.0in, 
bbllx=140pt,bblly=200pt, bburx=500pt, bbury=510pt} \\
\caption{{\it Left} -- pair fraction {\it vs.} string and association size ratio
diagram. {\it Right} -- average \dc value {\it vs.} string and association size ratio diagram.
Large magenta solid dots --- associations detected between T = 0 and 210 Kyr;
large magenta circles --- associations detected between T = 210 Kyr and 250 Kyr;
small magenta dots --- associations detected after 250 Kyr;
black crosses -- associations detected in the synthetic NEO populations;
red star --- current location of Drummond's (2000) A1 association;
blue open triangles --- current locations of Drummond's (2000) associations
A2 -- A14.
}
\label{fig.IdentifyingActualGeneticFamilies}
\end{figure}

In \S\ref{sec.DecoherenceOfSyntheticA1Family} we showed that
a genetic NEO family loses \dc coherence $\sim250$ Kyr after
creation. However, even after the family diffuses into the background
the volume search method still
detects associations with $>15$ objects
(Fig.~\ref{fig.SyntheticA1Evolution}B) but most of the members of these
associations are interlopers from the background. The question is then
how to distinguish between associations containing actual genetically
related members and those containing interlopers.

Comparing Fig.~\ref{fig.SyntheticA1Evolution}B with
Figs.~\ref{fig.SyntheticA1Evolution}C\&D reveals a clear difference
between the evolution of the association size and
the string size {\it or} total number of identified pairs --- after the family loses coherence the association size becomes very
noisy, jumping erratically between small and large families; while the
string size and number of pairs drops to a near steady-state
value with little variation from time step to time step. We utilize the difference in behaviour between associations, strings
and pairs to differentiate between real and false genetic NEO families in
Fig.~\ref{fig.IdentifyingActualGeneticFamilies}. In this simulation the strings are found using the members
of the association identified at each time step as seeds. The pairs are identified only amongst the members of the
association at each time step and the average \dc is also calculated
only amongst the same members.
Our intent is to mimic the fact that when searching for new genetic NEO families 
the members of the association will not be known {\it a priori} as in 
Figs.~\ref{fig.SyntheticA1Evolution} and \ref{fig.ActualA1Evolution}. In that figure the 
{\it pair fraction} is the total number of pairs identified within 
the association divided by $C_{n}^{2}=n(n-1)/2$, the total number of possible
unique pairs within the association when $n$ is the number of objects 
it contains.  The average \dc value is the sum of the 
\dc values between each
unique pair of objects in the association also divided by $C_{n}^{2}$. 
These two parameters essentially measure the ``tightness" of an association.  
We expect that fresh genetic NEO families will be very ``tight" with 
pair fractions approaching unity and average \dc value approaching
zero.  There is a very strong correlation between the pair fraction
and average \dc value.

The data in Fig.~\ref{fig.IdentifyingActualGeneticFamilies} are for the
synthetic A1 family of \S\ref{sec.DecoherenceOfSyntheticA1Family}.
Each data point represents a single time step in the dynamical
evolution of the synthetic A1 family with the different symbols representing
time since its formation.  There is clearly a strong time-dependent evolution 
that may be used to differentiate between real and false 
genetic NEO families. The 0.1 $D_{cutoff}$ is found epirically to
yield the clearest separation between associations detected at
different epochs.  The black crosses in the figure represent the
results of the same analysis applied to our synthetic NEO background
populations (see \S\ref{sec.StatisticalSignificanceOfNEOGroups}). 
The groupings found in those
populations (with no families) overlap with the ones found after the decoherence of the
synthetic A1 family but are well-separated from the groupings
found before the decoherence time of 250 Kyr.  The current location of
Drummond's (2000) A1 association is indicated in 
Fig.~\ref{fig.IdentifyingActualGeneticFamilies} with a red star, and the
other associations (A2 -- A14) are marked with blue triangles.

\section{Discussion}
\label{sec.Discussion}

We believe that the results described above bring into serious
question the reality of any of the NEO associations reported by
\citet{dru00}.  It is unlikely that a large NEO, similar in size to
the precursor to Drummond's (2000) A1 association, would have disrupted
during the last 300 Kyr, the time frame during which the family of
asteroids would be detectable with the \dc orbital similarity
technique.  The fact that the largest, and presumably most likely,
association reported by \citet{dru00} would be unidentifiable in just
a few hundred thousand years implies that, if it is real, it was found
just in the $\sim$250 Kyr period before the association becomes
totally decoherent (see Figure~\ref{fig.SyntheticA1Evolution}).  This
represents a mere $\sim$10\% of the dynamical lifetime of typical NEOs
and is in agreement with the decoherence time scales for NEO meteorite
streams as reported by \citet{pau05}.  Furthermore, the SFD
of associations detected in the actual NEO sample matches
the same distribution for associations detected in a random sample of
NEOs that accounts for observational selection effects.

Confusing the issue slightly, we applied our technique to all of
Drummond's (2000) original associations but using his more relaxed
($D_{cutoff}$) threshold on the \dc of 0.115 instead of our value of
0.1.  All but one of these associations remained firmly in the region
of random background fluctuations shown in
Fig.~\ref{fig.IdentifyingActualGeneticFamilies}.  Only when using the
original members of the A14 association as seeds with the relaxed \dc
threshold did we detect a 5-member string (as opposed to a 2-member
``string" with $D_{cutoff} = 0.1$) containing all four original A14
members (String Size/Association Size = 1.2).  At the same time, we
found 3 pairs amongst the four A14 members (\ie, Pair Fraction = 0.5)
compared to only 1 pair when $D_{cutoff} = 0.1$.  In this case, A14
would be located in the more interesting region of
Fig.~\ref{fig.IdentifyingActualGeneticFamilies} where genetic families
may be identified, well beyond the noise region defined by the fake
detections within the synthetic NEO populations. This hints that A14
may be a {\it genetic} NEO family but since it has only 4 members,
losing or gaining a single member in the string would cause a 25\%
difference in the ratio of the string size to association size.
Furthermore, our diagnostic technique is based on the simulation of a
25 member A1-like family rather than a 4 member A14-like family so the
comparison is not strictly applicable.

On the other hand, we show in
Section~\ref{sec.DecoherenceOfSyntheticA1Family} that a NEO family
{\it can} maintain coherence in it's orbital elements for $\gtrsim$200 Kyr.
So the \dc technique may be useful for identifying NEO
families that must then undergo further study before being classified
as real.  The supplementary studies should include at least a
statistical study of whether groups of the identified size are
unlikely in a similarly sized synthetic data set that
incorporates all the known observational selection effects.

Strong support for the veracity of a putative association found by the volume
search method may be obtained from its location in diagnostic diagrams similar to
Fig.~\ref{fig.IdentifyingActualGeneticFamilies}.  All the points with
pair fraction $\gtrsim 0.2$ and string size / association size
$\gtrsim 0.8$ correspond to positive detections of real genetic
associations with very few interlopers.
Alternatively, the figure may be considered as demonstrating the intrinsic
limits of \dc based methods in identifying families -- only the
tightest families (\ie, youngest ones) can be solidly detected. We note
that the current location of Drummond's (2000) A1 is solidly in the territory
occupied by random fluctuations.

We understand that we have only examined in detail a single association
identified by \citet{dru00}.  It is possible that parent-body NEOs in
other orbits may be more likely to disrupt and form longer living,
more easily identified NEO families.  It is also possible that one of
the other methods of producing an NEO family may produce longer lived
or more easily identifiable families.  The different production
mechanisms may even show a difference in the distribution of the
orbital elements of their members during the early stages of their
separated evolution.

For instance, the tidal disruption process \citep[\eg,][]{ric98} seems capable
of producing contact and close binary asteroids and also families of
objects on very similar orbits (\eg,\ Comet Shoemaker-Levy 9, \citet{bos94}; crater
chains on the various planets and their satellites, \citet{bot97}).  The fragments
from this production process may well be detectable as a family long
after the planetary close approach that produces the chain of
objects.  We plan on studying this possibility in future simulations.

The dynamical simulations of both the actual and synthetic A1 families
(\S\ref{sec.ActualA1Evolution} and
\S\ref{sec.DecoherenceOfSyntheticA1Family}) did not account for the
Yarkovsky effect \citep[see][for a review]{bot02b} that is known to
cause diffusion of orbital elements for small objects in the MB.  The
effect of this non-dynamical force will be correspondingly greater for
NEOs due to their approaching much closer to the Sun. Even so, the
expected drift rate for a 100 m diameter NEO with $a=1$ AU,
$i=45\arcdeg$ and typical density, albedo and surface conductivity, is
only $\sim 2\times 10^{-4}$ AU in $10^5$ years (Nesvorn{\' y}, personal
communication).  Thus, the Yarkovsky effect operates far too slowly to
contribute to the decoherence of NEO families.


It is clear that contemporary studies of genetic NEO families are
hampered by the small number of known NEOs thus, larger and deeper
surveys are needed. Pan-STARRS will be the next major survey to come
online and will detect as many asteroids and comets in one lunation as
are currently known.  Pan-STARRS will revolutionize our understanding
of NEOs --- while 3,319 NEOs are known as of April 2004 there are
hundreds of thousands of NEOs brighter than Pan-STARRS' expected
limit of $R \sim 24$ mag.  10,000 new NEOs should be discovered in the
first year of operations alone.  By the end of the first year of
Pan-STARRS operation, the completeness of NEOs of 1 km (100 m) in
diameter will be boosted to at least 90\% (20\%) from the current 75\%
(5\%), and after 10 years of operation Pan-STARRS will discover almost
all of the NEOs larger than 1 km diameter and more than half of the
NEOs bigger than 100 m diameter.  With a good understanding of the
observational selection effects in a single survey it should be
possible to identify NEO families or determine that there are none and
use this information to constrain the collision rate of NEOs.

\section{Conclusion}
\label{sec.Conclusion}

We find it unlikely that any of the enhancements in orbital
element space for the known NEO population 
are due to a genetic relationship
between the member objects.  The primary motivation for this
conclusion is the fact that synthetic data sets that incorporate
realistic distributions of NEOs and observational selection effects
show the same distribution of density enhancements as the actual
population.  We also base our conclusion on the fact that the current
understanding of NEO collision rates makes it extremely unlikely that
a large enough NEO could have disrupted recently enough to allow its
fragments to maintain enough coherence to remain detectable by the
\dc technique.  

We have identified a new technique that will allow future searches for
genetic families within the known NEO population.  The method relies
on the tight clustering in the \dc during the first couple hundred
thousand years after the production of the family.  Once an
association is identified, it is searched for strings and pairs.  If
the number of found pairs is more than $\sim$20\% of the maximum
possible number of pairs in the association {\it and} the ratio of the
maximum string size to association size is $\gtrsim 0.8$, then the
association is likely to be a real genetic family.

Upcoming improvements in NEO survey technology
such as Pan-STARRS may provide a large enough sample of NEOs in a
single, well-characterized survey, to allow identification of NEO
families or set a limit on their number.  This, in turn, will allow
dynamicists to refine their models of the collisional evolution of the
solar system.

\section{Acknowledgements}
Scotti's salary is supported by these grants to R. S. McMillan: NASA
Planetary Astronomy NNG04GK48G, NASA Near-Earth Object Observations
NAG5-13328, U. S. Air Force Office of Scientific Research
F49620-03-10107, and The Paul G. Allen Charitable Foundation.  We
thank Josh Barnes, Alberto Cellino, Jack Drummond, Nader Haghihipour,
Robert S. McMillan and David Nesvorn{\' y} for constructive suggestions
during the preparation of this manuscript.

\clearpage
 
\begin{table*}[!t]
\small
\center{
\begin{tabular}{llllcccc}
\hline
\hline
      &    &     &       &      & Effective  & Effective  & Effective  \\
Number&Name&Designation & H     & Tax. & Diameter (km)   & Diameter (km)   & Diameter (km)   \\      &    &     &       &      & ($p_V=0.306$) & ($p_V=0.128$) & ($p_V=0.070$) \\

\hline
1627 & Ivar&         & 13.2  & S    & 5.50       & 8.50       & 11.50      \\
5836 & & 1993 MF      & 13.9  & S    & 3.98       & 6.16       & 8.33       \\
2368 & Beltrovata&   & 15.21 & SQ   & 2.18       & 3.37       & 4.56       \\
3102 & Krok&         & 15.6  & S    & 1.82       & 2.81       & 3.81       \\
13553 & & 1992 JE    & 16.0  &      & 1.51       & 2.34       & 3.17       \\
12923 & & 1999 GK4   & 16.1  & S    & 1.45       & 2.24       & 3.02       \\
39796 & & 1997 TD    & 16.3  &      & 1.32       & 2.04       & 2.76       \\
& & 1998 KU2         & 16.61 & F,Cb & 1.14       & 1.77       & 2.39       \\
10860 & & 1995 LE    & 17.3  &      & 0.83       & 1.29       & 1.74       \\
8034 & Akka&         & 17.9  & S,Q  & 0.63       & 0.98       & 1.32       \\
27031 & & 1998 RO4   & 17.9  &      & 0.63       & 0.98       & 1.32       \\
65996 & & 1998 MX5   & 18.06 & X    & 0.59       & 0.91       & 1.23       \\
& & 1987 SF3         & 18.69 &      & 0.44       & 0.68       & 0.92       \\
8014 & & 1990 MF     & 18.7  &      & 0.44       & 0.68       & 0.91       \\
& & 1989 RC          & 18.75 &      & 0.43       & 0.66       & 0.89       \\
& & 1991 RJ2         & 19.15 &      & 0.36       & 0.55       & 0.74       \\
& & 1998 MR24        & 19.15 &      & 0.36       & 0.55       & 0.74       \\
26310 & & 1998 TX6   & 19.2  & C    & 0.35       & 0.54       & 0.73       \\
& & 1972 RB          & 19.24 &      & 0.34       & 0.53       & 0.71       \\
& & 1998 ME3         & 19.25 & F    & 0.34       & 0.52       & 0.71       \\
26817 & & 1987 QB    & 19.5  &      & 0.30       & 0.47       & 0.63       \\
& & 1997 RT          & 19.8  & O    & 0.26       & 0.41       & 0.55       \\
& & 1994 TA2         & 20.31 &      & 0.21       & 0.32       & 0.44       \\
& & 1998 QQ52        & 20.89 &      & 0.16       & 0.25       & 0.33       \\
& & 1995 SA4         & 22.26 &      & 0.08       & 0.13       & 0.18       \\
\hline
 & &  &       & Lower limit $\rightarrow$     & 6.37 km    & 9.84 km    & 13.31 km   \\
\hline

\end{tabular}
} 


\caption{\small The A1 association of \citet{dru00} in order of increasing
  absolute magnitude ($H$).  This is the largest
  association identified within the NEO population known at the time.
  Our nominal values for the assumed diameters of the object
  correspond to the intermediate albedo value of $p_V=0.128$.}

\label{tab.A1Association}

\end{table*}

\begin{table}[!t]
\small
\center{
\begin{tabular}{ccccccc}
\hline
\hline
Grouping   &          & &  Association     &          & &   \\
(Fu \etal) &  Members & &  \citep{dru00} & Members  & & Overlap  \\
\hline
     1     &    29    & &        A1        &   25     & &    25    \\
     2     &    23    & &        A8        &   11     & &    11    \\
     3     &    23    & &        A4        &   10     & &     9    \\
     4     &    22    & &        A2        &   15     & &    15    \\
\hline

\end{tabular}
} 
\caption{A comparison between the four largest associations identified in
  this analysis and in \citet{dru00}.  The four largest groups
  found in this analysis contain four of the associations identified
  by \citet{dru00}.}

\label{tab.LargeGroupComparison}

\end{table}


\clearpage

\clearpage

\clearpage

\clearpage

\clearpage

\clearpage

\clearpage

\end{document}